\documentstyle[sprocl,epsf]{article}

\renewcommand{\bar}[1]{\overline{#1}}
\newcommand{\longvec}[1]{\overrightarrow{\!\!#1}}
\newcommand{\VEV}[1]{\left\langle{#1}\right\rangle}

\newcommand{\ket}[1]{\,\left|\,{#1}\right\rangle}
\newcommand{\M}{{\cal M}}
\newcommand{\R}{{\cal R}}
\newcommand{\etal}{{\em et al.}}
\newcommand{\ie}{{\em i.e.}}
\newcommand{\eg}{{\em e.g.}}

\renewcommand{\bar}[1]{\overline{#1}}
\newcommand {\dbar}{\bar d}
\newcommand {\qbar}{\bar q}
\newcommand {\cbar}{\bar c}

\newcommand{\gsim} {\buildrel > \over {_\sim}}

\begin{document}

\title{Light-Cone Quantized QCD and Novel Hadron Phenomenology}

\author{ S. J. Brodsky }

\address{Stanford Linear Accelerator Center\\ Stanford University, Stanford,
California 94309}

\maketitle\abstracts{
I review progress made in solving gauge theories such as collinear quantum
chromodynamics using light-cone Hamiltonian methods.
I also show how the light-cone Fock expansion for hadron wavefunctions can be
used to compute operator matrix elements such as decay amplitudes, form factors,
distribution amplitudes, and structure functions, and how it provides a tool for
exploring novel features of QCD.  I also review commensurate scale
relations, leading-twist identities which relate physical observables to each
other, thus eliminating renormalization scale
and scheme ambiguities in perturbative QCD predictions.} 

\section{Introduction}

The key challenge of nonperturbative quantum chromodynamics is to
compute the
spectrum of hadrons and gluonic states from first principles as well as
determine the wavefunctions for each QCD bound state in terms of its quark and
gluon degrees of freedom.  If we had such a complete solution, then we
could compute the quark and gluon structure functions and distribution
amplitudes
which control hard-scattering
inclusive and exclusive reactions, as well as all of the operator matrix
elements of currents which underlie electro-weak form factors and the weak decay
amplitudes of the light and heavy hadrons.  The knowledge of hadron
wavefunctions
would also provide a deep understanding of the physics of QCD at the
amplitude level, illuminating exotic effects of the theory such as color
transparency, intrinsic heavy quark effects, hidden color, diffractive
processes, and the QCD van der Waals interactions.

Solving a quantum field theory such as QCD is clearly not easy.  However, highly
non-trivial, one-space one-time relativistic quantum field theories which
mimic many
of the features of QCD  have already been completely solved using light-cone
Hamiltonian methods.\cite{PinskyPauli}  In fact, virtually any
(1+1) quantum field theory can be solved using the method of Discretized
Light-Cone-Quantization (DLCQ).
\cite{DLCQ,Schlad}  In DLCQ, a quantum field theory is
rendered discrete in momentum space by imposing periodic or anti-periodic
boundary conditions.  The Hamiltonian
$H_{LC}$, which can be constructed from the Lagrangian using light-cone time
quantization, can then be diagonalized, in analogy to Heisenberg's
solution of the eigenvalue problem in quantum mechanics.  In the one-space
one-time theories, the diagonalization is a straightforward computational
problem, and the resulting eigenvalues and eigensolutions then provide the
complete spectrum of hadrons, together with their respective light-cone
wavefunctions.

A beautiful illustration of the application of light-cone quantization
to the solution of a quantum field theory is the DLCQ analysis of ``collinear"
QCD: a variant of $QCD(3+1)$ defined by
dropping all of interaction terms in $H^{QCD}_{LC}$ involving
transverse momenta.\cite{Kleb} Even though this theory is effectively
two-dimensional, the transversely-polarized degrees of freedom of the gluon
field
are retained as two scalar fields.  Antonuccio and Dalley
\cite{AD} have recently used DLCQ to solve this theory.  The diagonalization of
$H^{\rm collinear}_{LC}$ provides not only the complete bound and continuum
spectrum
of the collinear theory, but it also yields the
complete ensemble of light-cone Fock state wavefunctions needed to construct the
quark and gluon structure functions of each hadronic and gluonic state.
For example,  Antonuccio and Dalley
obtain the spectrum of gluonia, and the polarized gluon and quark structure
functions of the mesonic states.
Although the
collinear theory is a drastic approximation to physical $QCD(3+1)$, the
phenomenology of its DLCQ solutions demonstrate general features of gauge
theory, such as the peaking of
the wavefunctions at minimal invariant mass, color
coherence, and the helicity retention of leading partons in the polarized
structure functions at $x\rightarrow 1$.
The DLCQ solutions of the one-space one-time gauge theories can be
obtained for arbitrary coupling strength, flavors, and
colors.

The solutions to collinear QCD provide a ``standard candle" or
theoretical laboratory in which other nonperturbative methods proposed to solve
QCD, such as lattice gauge theory, Bethe-Salpeter methods,  and various
approximations can be tested and compared.

The fact that one actually solve a
non-trivial relativistic quantum field theory in one space and one
time gives hope
that the full solutions to QCD(3+1) will eventually be accomplished.  In these
lectures I shall also discuss the possibility
that one can use the collinear theory as a first approximation to a procedure
which systematically constructs the full wavefunction solutions of QCD(3+1).
I will also outline the progress made in understanding
hadrons at the amplitude level, using the light-cone Fock expansion as a
physics tool for
exploring novel features of QCD in hadron physics.  I also review
commensurate scale relations, a method which relates physical observables to
each other, thus eliminating ambiguities due to scale and scheme ambiguities.

\section{The Light-Cone Fock Expansion}  

The concept of the ``number of constituents" of a
relativistic bound state such as a hadron in quantum
chromodynamics, is not only frame-dependent, but its value can
fluctuate to an arbitrary number of quanta. Thus when a laser beam crosses a
proton at fixed ``light-cone" time
$\tau = t+z/c= x^0 + x^z$, an interacting photon can
encounter a state with any given number of quarks, anti-quarks, and
gluons in flight (as long as $n_q - n_{\bar q} = 3$). The probability amplitude
for each such 
$n$-particle state of on-mass shell quarks and gluons in a hadron is given by a
light-cone Fock state wavefunction
$\psi_{n/H}(x_i,\vec k_{\perp i},\lambda_i)$, where the constituents have
positive
longitudinal light-cone momentum fractions
\begin{equation}
x_i = \frac{k^+_i}{P^+} = \frac{k^0+k^z_i}{P^0+P^z}\ ,\quad \sum^n_{i=1} x_i= 1
\ ,
\label{eq:c}
\end{equation}
relative transverse momentum
\begin{equation}
\vec k_{\perp i} \ , \quad \sum^n_{i=1}\vec k_{\perp i} = \vec 0_\perp \ ,
\label{eq:d}
\end{equation}
and helicities $\lambda_i.$ 
The ensemble
\{$\psi_{n/H}$\} of such light-cone Fock
wavefunctions is a key concept for hadron physics, providing a conceptual basis
for representing physical hadrons (and also nuclei) in terms of their fundamental
quark and gluon degrees of freedom.\cite{BR}  In the light-cone formalism, 
the vacuum is essentially trivial.    
Since each particle moves forward in light-cone time $\tau$ 
with positive light-cone momenta fractions $x_i$,  
light-cone perturbation theory is particularly simple and intuitive, 
involving many fewer diagrams than equal-time theory.  

The light-cone Fock expansion is defined in the following way:  one
first constructs the light-cone time evolution operator $P^-=P^0-P^z$
and the invariant mass operator $H_{LC}= P^- P^+-P^2_\perp $ in 
light-cone gauge $A^+=0$ from the QCD Lagrangian.
The total longitudinal momentum $P^+ = P^0 + P^z$ and transverse
momenta $\vec P_\perp$ are conserved, \ie, are independent of the interactions.
The matrix elements of
$H_{LC}$ on the complete orthonormal basis $\{\vert n >\}$ 
of the free theory $H^0_{LC} =
H_{LC}(g=0)$ can then be constructed.  The matrix elements
$\VEV{n\,|\,H_{LC}\,|\,m}$ connect Fock states differing by 0,
1, or 2 quark or gluon quanta, and they include the instantaneous quark
and gluon contributions imposed by eliminating dependent degrees of
freedom in light-cone gauge.  

In practice it is essential to introduce an
ultraviolet regulator in order to limit the total range of
$\VEV{n\,|\,H_{LC}\,|\,m}$, such as a ``global" cutoff in the invariant
mass of the free Fock states: 
\begin{equation} \M^2_n = \sum^n_{i=1}
\frac{k^2_\perp + m^2}{x} < \Lambda^2_{\rm global} \ . \label{eq:a}
\end{equation} 
One can also introduce a ``local" cutoff to limit
the change in invariant mass $|\M^2_n-\M^2_m| < \Lambda^2_{\rm local}$
which provides spectator-independent regularization of the
sub-divergences associated with mass and coupling renormalization. 

The
natural renormalization scheme for the coupling is $\alpha_V(Q)$, the
effective charge defined from the scattering of two infinitely heavy
quark test charges.  The renormalization scale can then be determined
from the virtuality of the exchanged momentum, as in the BLM and
commensurate scale methods.\cite{BLM,BGKL} I will 
discuss this further in Section 18.

In the DLCQ method , the matrix
elements 
$\VEV{n\,|\,H^{(\Lambda)}_{LC}\,|\,m}$, are made discrete in momentum space by
imposing periodic or anti-periodic boundary conditions in $x^-=x^0 - x^z$ and
$\vec x_\perp$ (see Section 3). Upon diagonalization of $H_{LC}$, the eigenvalues provide the
invariant mass of the bound states and eigenstates of the continuum.  The
projection of the hadronic eigensolutions on the free Fock basis define the
light-cone wavefunctions. For example, for the proton, 
\begin{eqnarray}
\ket p &=& \sum_n \VEV{n\,|\,p}\, \ket n \nonumber \\
&=& \psi^{(\Lambda)}_{3q/p} (x_,\vec k_{\perp i},\lambda_i)\,
\ket{uud} \\[1ex]
&&+ \psi^{(\Lambda)}_{3qg/p}(x_i,\vec k_{\perp i},\lambda_i)\,
\ket{uudg} + \cdots \nonumber
\label{eq:b}
\end{eqnarray}
The light-cone formalism has the remarkable feature that the 
$\psi^{(\Lambda)}_{n/H}(x_i,
\vec k_{\perp i},\lambda_c)$ are invariant under longitudinal boosts; \ie,\ they
are independent of the total momentum $P^+$, $\vec P_\perp$ of the
hadron.  As we shall discuss below, given the
$\psi^{(\Lambda)}_{n/H},$ we can construct any electromagnetic or electroweak form
factor from the diagonal overlap of the LC wavefunctions.\cite{BD} 
This is illustrated in detail in Section 6.
Similarly,
the matrix elements of the currents that define quark and gluon
structure functions can be computed from the integrated squares of the LC
wavefunctions.\cite{BrodskyLepage}

These properties of the LC formalism are all highly-nontrivial features. 
In contrast, in equal-time formalism, the evaluation of any electromagnetic form
factor requires the computation of non-diagonal matrix elements of bound state
Fock wavefunctions in which the parton number can change by two units; even worse,
matrix elements involving spontaneous pair production or annihilation is
also required, so a complete solution of the vacuum is also required.  
In light-cone quantization, the full vacuum is also the vacuum of the free theory and thus is trivial.   Further,
the equal-time computation is only valid in one Lorentz frame; boosting the result to a
different frame is a dynamical problem as complicated as solving the complete
Hamiltonian problem itself.

In general, any hadronic amplitude such as quarkonium decay, heavy hadron
decay, or any hard exclusive hadron process can be constructed as the
convolution of the
light-cone Fock state wavefunctions with quark-gluon matrix elements 
\cite{BrodskyLepage}
\begin{eqnarray}
\M_{\rm Hadron} &=& \prod_H \sum_n \int
\prod^{n}_{i=1} d^2k_\perp \prod^{n}_{i=1}dx\, \delta
\left(1-\sum^n_{i=1}x_i\right)\, \delta
\left(\sum^n_{i=1} \vec k_{\perp i}\right) \nonumber \\[2ex]
&& \times \psi^{(\Lambda)}_{n/H} (x_i,\vec k_{\perp i},\Lambda_i)\, \M
^{(\Lambda)}_{q,g} \ . 
\label{eq:e}
\end{eqnarray}
Here $\M^{(\Lambda)}_{q,g}$ is the underlying quark-gluon
subprocess scattering amplitude, where the (incident or final) hadrons are
replaced by quarks and gluons with momenta $x_ip^+$, $x_i\vec
p_{\perp}+\vec k_{\perp i}$ and invariant mass above the 
separation scale $\M^2_n > \Lambda^2$. The LC ultraviolet regulators thus
provide a  {\it LC factorization scheme} for elastic and inelastic
scattering, separating the hard dynamical contributions with invariant mass
squared $\M^2 >
\Lambda^2_{\rm global}$ from the soft physics with $\M^2 \le \Lambda^2_{\rm
global}$ which is incorporated in the nonperturbative LC wavefunctions.  The
DGLAP evolution of parton distributions can be derived 
by computing the variation of the Fock expansion with respect to
$\Lambda^2_{\rm global}$.\cite {BrodskyLepage}

The use of the global cutoff to separate hard and soft physics is more than a
convention; it is essential in order to correctly analyze the behavior of
deep inelastic scattering structure functions in the $x_{bj} \to 1$ endpoint
regime.\cite{BrodskyLepage}  At large
$x$, the spectator constituents of the hadron target are forced to stop,
placing the struck quark far off shell. From the stand-point of the LC Hamiltonian
theory, the LC energy $M^2 - {\cal M}_n^2$ becomes infinitely negative. 
Similarly, in the covariant formalism, the Feynman virtuality becomes infinitely
spacelike:
$k_F^2-m_q^2 = (p-p_s)^2-m_s^2 = - x( M^2 - {\cal M}_n^2)  \to
-(m_s^2+k^2_\perp)/(1-x)$ for
$x \to 1$. Here $x$ is the light-cone momentum fraction of the light quark, $p_s$
is the four-momentum of the remnant spectator system with  $p^2_{s} = m^2_s > 0.$
Thus in the large $x$ regime, where $k_F^2$ becomes far off-shell, one cannot separate
the hard physics from the physics of the target wavefunction.  The LC
factorization scheme correctly isolates this phenomena.  An important physical
consequence is that DGLAP evolution is truncated; effectively, the starting
evolution scale is of order $-k^2_F$. Thus in the $x_{Bj} \sim 1$ regime where
$W^2 = (p+q)^2 = (1-x_{Bj}) Q^2/x_{bj}$ is fixed, the DGLAP evolution due to the
radiation of hard gluons is truncated and the power law behavior
$(1-x_{Bj})^n$  dictated by the underlying hadron wavefunctions is maintained.  It
is this fact which allows exclusive-inclusive duality in deep
inelastic lepton scattering to be maintained: the nominal power law (dimensional
counting) behavior of structure functions at $x
\to 1$ and the power-law fall off of form factors at large $Q^2$  match at fixed
$W^2$.

The simplest, but most fundamental, characteristic of a hadron in the light-cone
representation, is the hadronic distribution amplitude,\cite{BrodskyLepage} defined
as the integral over transverse momenta of the valence (lowest particle
number) Fock wavefunction; \eg, for the pion
\begin{equation}
\phi_\pi (x_i,Q) \equiv \int d^2k_\perp\, \psi^{(Q)}_{q\bar q/\pi}
(x_i, \vec k_{\perp i},\lambda)
\label{eq:f}
\end{equation}
where the global cutoff $\Lambda_{global}$ is identified with the
resolution $Q$.  The distribution amplitude controls leading-twist exclusive
amplitudes at high momentum transfer, and it can be related to the gauge-invariant
Bethe-Salpeter wavefunction at equal light-cone time
$\tau = x^+$.  The distribution amplitude  is boost and gauge
invariant.  Its  $\log Q$  evolution can be derived from the
perturbatively-computable tail of the valence light-cone wavefunction in the
high transverse momentum regime.\cite{BrodskyLepage,EfremovRad} 

Exclusive processes are particularly challenging to compute in quantum
chromodynamics because of their sensitivity to the unknown
nonperturbative bound state dynamics of the hadrons. However, in some
important cases, the leading power-law behavior of an exclusive
amplitude at large momentum transfer can be computed rigorously in the
form of a factorization theorem which separates the soft and hard
dynamics. For example, the leading $1/Q^2$ fall-off of the meson form
factors can be computed as a perturbative expansion in the QCD
coupling\cite{BrodskyLepage,EfremovRad}:
\begin{equation}
F_M(Q^2)=\int^1_0 dx \int^1_0 dy
\phi_M(x, {\tilde Q}) T_H(x,y,Q^2)
\phi_M(y,{\tilde Q}),
\end{equation}
where $\phi_M(x,{\tilde Q})$ is the process-independent meson
distribution amplitude which encodes the nonperturbative dynamics of
the bound valence Fock state up to the resolution scale ${\tilde Q}$,
and
\begin{equation}
T_H(x,y,Q^2) = 16 \pi C_F {\alpha_s(\mu)\over (1-x) (1-y)
Q^2}\left(1 + O(\alpha_s)\right)
\end {equation}
is the leading-twist perturbatively calculable subprocess amplitude
\hfill\break
$\gamma^* q(x) \overline q(1-x) \to q(y) \overline q(1-y)$, obtained
by replacing the incident and final mesons by valence quarks collinear
up to the resolution scale ${\tilde Q}$.  The contributions from
non-valence Fock states and the correction from neglecting the
transverse momentum in the subprocess amplitude from the
nonperturbative region are higher twist, \ie, they are power-law
suppressed. The transverse momenta in the perturbative domain lead to
the evolution of the distribution amplitude and to
next-to-leading-order (NLO) corrections in $\alpha_s$.  The
contribution from the endpoint regions of integration, $x \sim 1$ and
$y \sim 1,$ are power-law and Sudakov suppressed and thus can only
contribute corrections at higher order in $1/Q$.\cite{BrodskyLepage}

The physical pion form factor must be independent of the separation
scale $\tilde Q.$  Again, it should be emphasized that the natural variable to
make this separation is the light-cone energy, or equivalently the invariant mass
${\cal M}^2 ={\vec{k_\perp}^2/ x(1-x)}$, of the off-shell 
partonic system.\cite{JIPang,BrodskyLepage}
Any residual dependence on the choice of
$\tilde Q$ for the distribution amplitude will be compensated by a
corresponding dependence of the NLO correction in $T_H.$ However, the
NLO prediction for the pion form factor depends strongly on the form
of the pion distribution amplitude as well as the choice of
renormalization scale $\mu$ and scheme. I will discuss recent progress in
eliminating such scheme and scale ambiguities and testing QCD in exclusive
processes in Sections 18 and 19.

\begin{figure}[htbp] 
\begin{center}
\leavevmode
\epsfbox{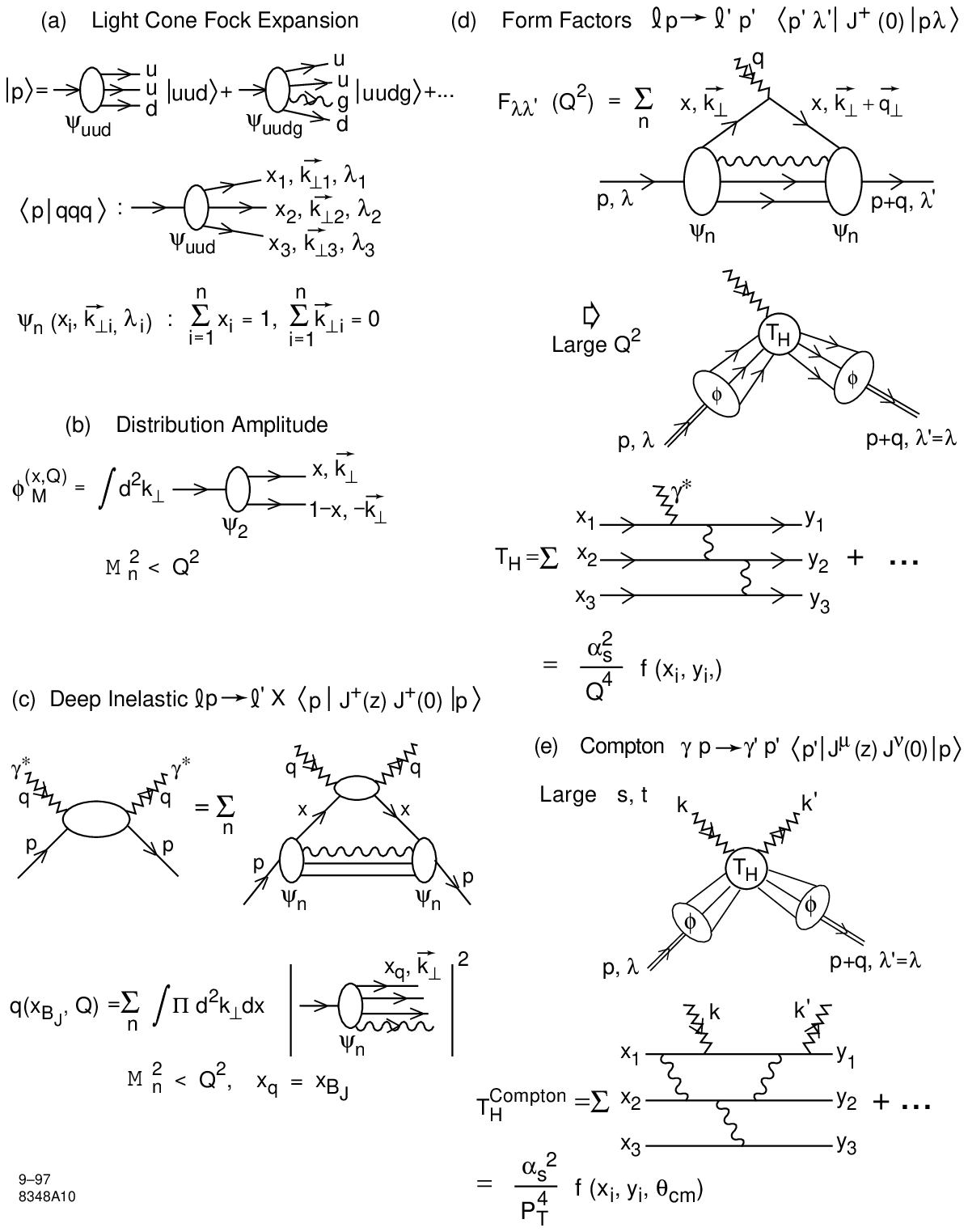}
\end{center}
\end{figure}

\begin{figure}[htbp]
\begin{center}
\leavevmode
\epsfbox{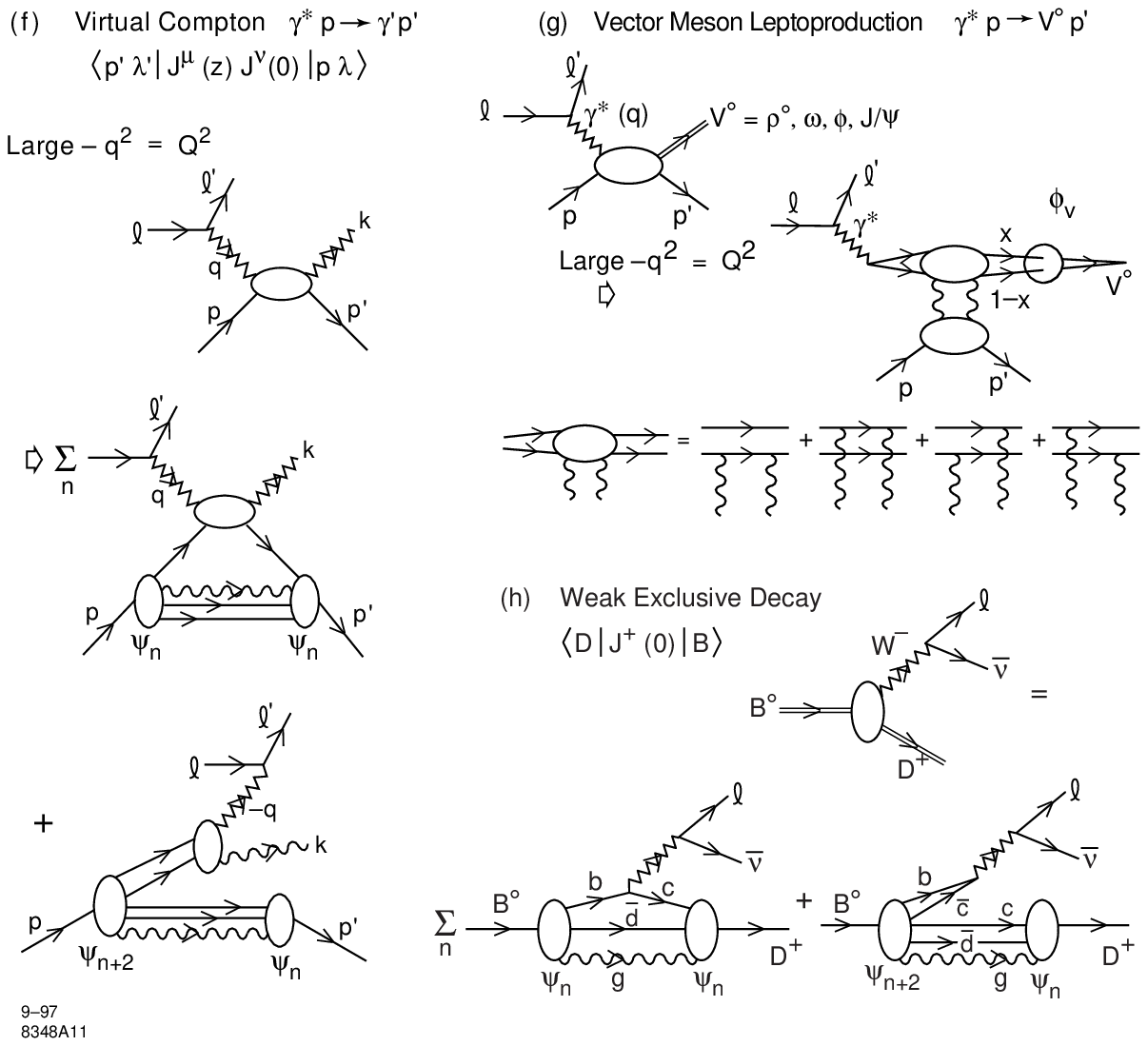}
\end{center}
\caption{Computation of QCD processes in terms of light-cone Fock wavefunctions.  
The various processes are described in the text.}
\label{fig:X}
\end{figure}      

A catalog of applications of light-cone Fock wavefunctions to QCD processes 
is illustrated in Fig. \ref{fig:X}.   
The light-cone expansion for a proton in 
terms of the complete set of color-singlet baryon number $B = 1$ 
free Fock states is illustrated in Fig. \ref{fig:X}a.  
The distribution amplitude which controls high momentum 
transfer mesonic processes is illustrated in Fig. \ref{fig:X}b.  
The meson distribution amplitude $\phi_M(x,Q)$ is defined such 
that the invariant mass ${\cal M}$ of the free partons in any 
intermediate state are cutoff at the ultraviolet scale $Q$.   
The relation of structure functions in deep inelastic lepton 
scattering to the integrated square of light-cone Fock
 wavefunctions is illustrated in Fig. \ref{fig:X}c.    
As shown in Fig. \ref{fig:X}d, the light-cone wavefunctions 
provide an exact basis for the computation of the matrix elements 
of spacelike local currents in terms of diagonal  overlap integrals 
with $x_i=x_i'$ unchanged and  parton number $n = n'$.  
Figure \ref{fig:X}d also illustrates the factorization of the 
nucleon form factors at high momentum transfer in terms of 
the convolution of a hard-scattering amplitude $T_H(x_i, y_i, Q)$ 
(for the scattering the valence quarks from the initial to final direction) 
with the nucleon distribution amplitudes.   
Figure \ref{fig:X}e illustrates the application of 
perturbative QCD factorization to the Compton scattering amplitude 
at large $-t$ and $s$.  
Figure \ref{fig:X}f illustrates the computation of a non-forward 
matrix element of currents; specifically virtual Compton scattering 
where the incident photon has large incident virtuality $q^2 = -Q^2$.   
Non-diagonal $n'= n-2$ Fock state convolutions are required when evaluating 
such processes in a collinear reference frame.  
The computation of a weak decay matrix element 
$B^0 \to D^+ \ell \bar \nu$ in terms of light-cone Fock wavefunctions is illustrated in 
Fig. \ref{fig:X}g.   
Both diagonal and non-diagonal overlap integrals contribute.  
Finally, the perturbative QCD factorization of the dominant contribution 
to vector meson leptoproduction at large photon virtuality $Q^2$ 
and high energy $s \gg -t, Q^2$ is illustrated in Fig. \ref{fig:X}h.  
The dominant contribution arises from longitudinally polarized photons, 
and the $q \bar q$ pair typically has small transverse size $b_\perp \sim 1/Q$.  
The coupling of the quark pair to the outgoing vector meson 
is controlled by the vector meson distribution amplitude.  

\section{DLCQ:  A Program for Solving QCD (3+1).}

The DLCQ method \cite{DLCQ} consists of diagonalizing the light-cone Hamiltonian at fixed
$x^+$ on a free Fock basis $\{\ket n\}$; \ie\ the complete set of eigenstates of
the free Hamiltonian $H^0_{LC}$ satisfying periodic or anti-periodic boundary
conditions in $x^-$.  The eigenvalue problem is
\begin{eqnarray} H_{LC} \ket\Psi &=& M^2\ket\Psi \\
\VEV{n\,|\,H_{LC}\, |\,m}\VEV{m\,|\,\Psi} &=& M^2\VEV{n\,|\,\Psi}
\end{eqnarray} with
\begin{equation} k^+_i = \frac{2\pi}{L}\, n_i > 0 \qquad P^+ = \frac{2\pi}{L}\,
K \qquad
\Sigma n_i = K \ .
\end{equation} Here $K$, the ``harmonic resolution,'' is an arbitrary positive
integer.  The continuum limit corresponds to $K\Rightarrow \infty$.  The value
of length $L$ is an irrelevant boost parameter in that it never appears in
physics results.  Since there are only a finite number of partitions of a given
$K$ among the positive integers $n_i$ with $\Sigma n_i = K$, the number of
distribution Fock states are automatically rendered discrete.  The transverse
momenta are also made discrete by choosing periodic or anti-periodic conditions
in $x_\perp$.  Then
\begin{equation}
\vec k_{\perp i} = \frac{2\pi}{L_\perp}\ \vec n_{\perp i} \ .
\end{equation} The limit on the number of states is then controlled by the
global cutoff.

The diagonalization of the light-cone Hamiltonian thus becomes the problem of
diagonalizing large Hermitian matrices, a numerical analysis problem, solvable
by Lanczos or other methods.  In the case of 1+1 dimensions, the problem is
completely tractable, so virtually any 1+1 quantum field theory can be solved in
this manner. 

In the case of QCD (1+1) in $A^+=0$ gauge, where there are no dynamical
gluons, the only interaction terms arise from ``instantaneous'' gluon exchange:
\begin{equation} H^{LC}_I = \frac{g^2}{\pi} \left[\frac{1}{(k^+-\ell^+)^2}
- \frac{1}{(k^++m^+)^2}\right]
\end{equation} corresponding to $t$-channel or $s$-channel contributions in the
amplitude
\begin{equation}
\bar q(k^+)\, q(m^+) + \bar q(\ell^+)\, q(n^+) \ .
\end{equation} There is also a mass renormalization contribution to $q(n^+)$
generated from normal ordering
\begin{equation}
\delta H = \frac{g^2}{\pi}\ \sum^{n^+}_{m=1}\
\frac{1}{m^2} \ .
\end{equation}

The solution to the diagonalization of the light-cone Hamiltonian produces not
only the mass eigenvalues of the theory, but also the eigensolutions, as
wavefunction coefficients in the LC Fock basis:
\begin{equation}
\ket\Psi = \sum_n \psi_n (x_i,\vec k_{\perp i},\lambda_i)\ket n
\end{equation} where each Fock component $\ket n$ has the same global and
conserved quantum numbers as the eigenstate.  The values of the light-cone
momentum fractions are evaluated at 
\begin{equation} x^+_i = \frac{k^+_i}{p^+} = \frac{n_i}{K} =
\left\{\frac{1}{K}\,, \frac{2}{K}\,, 
\frac{3}{K}\,\cdots \frac{K-1}{K}\right\} \ 
\end{equation} 
Thus one samples the wavefunctions at rational points which
approach the continuum theory at $K\rightarrow \infty$.  The absence of the
end-points at $x_i=0,1$ corresponds to the neglect of zero modes.  Except
for massless, collinear $k_1$, such parton configurations are associated with
infinitely massive free energy:
\begin{equation}
\M^2_n = \sum^n_{i=1}\,
\left(\frac{k^2_\perp+m^2}{x}\right)_i \rightarrow \infty
\end{equation} and thus exceed the global cutoff limit.  Physically, the $x
\rightarrow 0$ limit is associated with partons infinitely far in rapidity from
the center of mass of the bound state itself
\begin{equation} y_i - Y = \ell n\, x_i\ ;
\end{equation} such partons are only relevant at very large energies
$W^2=(p+q)^2$ in the computation of structure functions.  On the other hand, the
LC Fock wavefunctions do not necessarily vanish at $x_i=0$ since they may
correspond to soft gluons with $m^2$ and $k^2_\perp = 0$.
In general, even the fermion distribution need not vanish at $x \rightarrow 0$ 
in gauge theory since only the combination from the sum over states,
\begin{equation}
\sum \left[\frac{(\vec k_\perp-g\, \vec A_\perp)^2+m^2}{x}\right]_i
\end{equation} has to be finite in the interacting theory.  As Antonuccio and
Dalley and I \cite{ABD}\ have recently shown, the cancellation of infinities at
$x_i
\rightarrow 0$ for fermions in gauge theory imposes strict ``ladder relations''
between Fock states with one or two more or less gluons in the bound state.  We
have also shown how this type of analysis leads to Regge power-law behavior of
the quark distributions at $x \rightarrow 0$.

\begin{figure}[htbp] 
\begin{center}
\leavevmode
\epsfbox{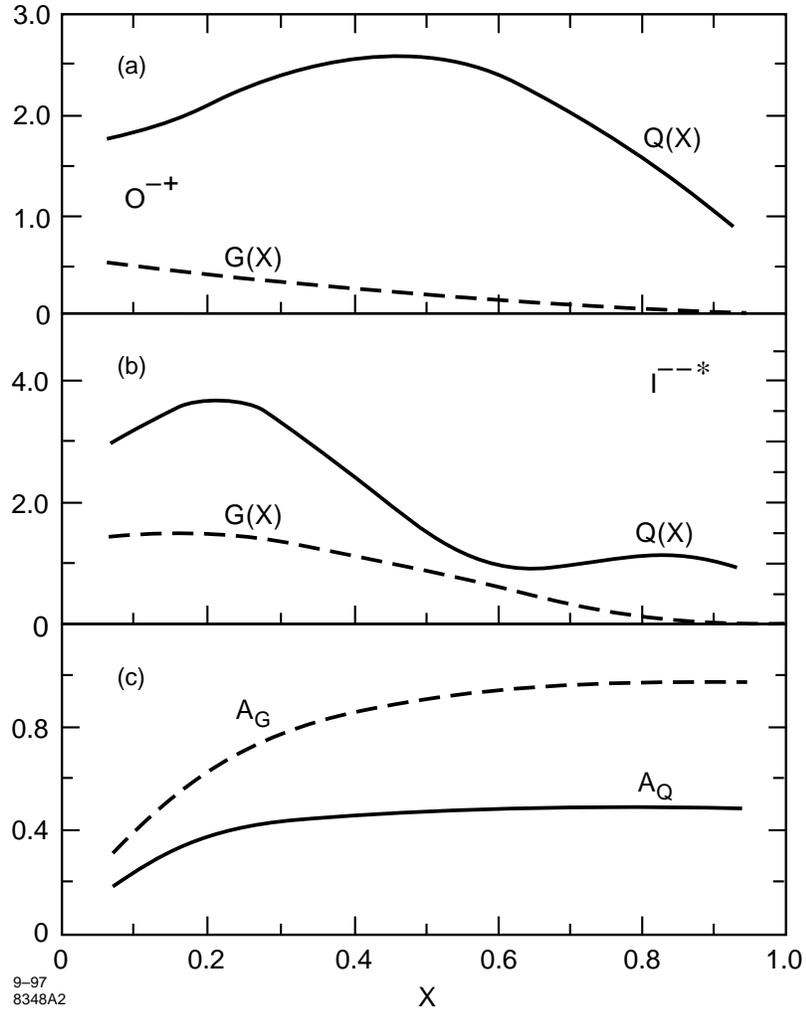}
\end{center}
\caption[*]{Examples of quark and gluon distributions in meson bound states 
in collinear QCD.  From Antonuccio and Dalley.\cite{AD} (a) Lowest $J_z=0$ mesonic state; (b) excited $J_z=1$ mesonic state; (c) longitudinal helicity asymmetry in lowest $J_z=1$ mesonic state.}
\label{fig:1}
\end{figure}      

These interesting features are illustrated in Fig. \ref{fig:1}a and \ref{fig:1}b 
which shows the
gluon $G(x)$ and quark $Q(x)$ distributions of the lowest $0^{-+}$ and first
excited $1^{--}$ mesonic bound states of collinear QCD $(N_c \rightarrow
\infty)$ from DLCQ.\cite{AD} Neither the quark nor the gluon distributions vanish at
$x\rightarrow 0$.  The polarization asymmetry for quarks and gluons with
helicity aligned and anti-aligned to the $J_z$ spin of $1^{--}$ ground state
shown in Fig. \ref{fig:1}c\ demonstrates the tendency toward  helicity alignment at $x
\rightarrow 1$.

\section{Jet Hadronization in Light-Cone QCD}

One of the goals of nonperturbative analysis in QCD is to compute jet
hadronization from first principles.  The DLCQ solutions provide a possible
method to accomplish this.  By inverting the DLCQ solutions, we can write the
``bare'' quark state of the free theory as
\begin{equation}
\ket{q_0} = \sum \ket n \VEV{n\,|\,q_0}
\end{equation} where now $\{\ket n\}$ are the exact DLCQ eigenstates of
$H_{LC}$, and 
$\VEV{n\,|\,q_0}$ are the DLCQ projections of the eigensolutions.  The expansion
in automatically infrared and ultraviolet regulated if we impose global cutoffs
on the DLCQ basis:
\begin{equation}
\lambda^2 < \Delta\M^2_n < \Lambda^2 \ .
\end{equation} where $\Delta\M^2_n = \M^2_n-(\Sigma \M_i)^2$.  It would be
interesting to study this type of jet hadronization at the amplitude level for
the existing DLCQ solutions to QCD (1+1) and collinear QCD.

\section{Light-Cone Quantization and Renormalization Theory}

The renormalization procedure for LC Hamiltonian theory is well understood in
perturbation theory.  For example, mass and coupling renormalization counter
terms can be introduced in the standard way in QED to absorb the ultraviolet
divergences at each order of perturbation theory.  An explicit method,
``alternating denominators", which provides an automatic method to construct the
local counter terms, was used\cite{BRS}\ to compute the lepton anomalous moments 
through order $\alpha^2$
and partly through order $\alpha^3.$  Lepage and I have employed an ultraviolet
Hamiltonian renormalization scheme to derive the hard-scattering expansion for
exclusive processes in QCD, including the evolution equations for the renormalized
distribution amplitudes.  Burkardt and Langnau\cite{refaa}\ have shown
that kinetic and vertex mass renormalization counter terms are needed  to 
restore the full invariance structure of the theory when one uses a light-cone 
regulation such as the global cutoff.  Burkardt\cite{BURKA} also has shown that tadpole 
diagram renormalization of $g\phi^4$ theory is consistently handled in the LC theory 
as a zero mode component to the renormalization of the scalar particles.  The renormalization of Yukawa theories has also been analyzed.\cite{BURKB} 
Hiller, McCartor, and I are working on a program in which the full Lorentz-invariant 
structure of light-cone Hamiltonian theory is restored using generalized Pauli-Villars
regularization.

One of the advantages of DLCQ is that it provides a convenient infrared
regularization of zero modes since they become discrete entities.  In some model
field theories, the zero modes take the place of the vacuum in reproducing the
physics of spontaneous symmetry breaking.  In other cases, such as the massive
Schwinger model QED (1+1), the zero modes allow a simulation of external
electric fields.  As yet it is not clear whether LC zero roles play an
essential role in analyzing QCD(3+1).

\section{Form Factors and Light-Cone Wavefunctions}

A critical advantage of the light-cone formalism is that the knowledge of the LC
Fock wavefunction is sufficient to compute the elastic electroweak form factors.
It is remarkable that all such matrix elements can be computed from diagonal
(parton-conserving) overlap integrals of the LC Fock wavefunctions. 
In this section I will review the light-cone formalism 
for the computation of form factors for both
elementary and composite systems.\cite{r12,r13,BD}   We can choose
light-cone coordinates with the incident lepton directed along the
$z$ direction\cite{r14} $(p^\pm\equiv p^0\pm p^3)$:
\begin{equation}
p^\mu \equiv (p^+,p^-,\longvec p_1) =
\left(p^+,\frac{M^2}{p^+},\longvec 0\!\!_\perp\right)\ ,\qquad
q = \left(0,\frac{2q\cdot p}{p^+}, \longvec q\!\!_\perp\right)\ ,
\label{eq6}
\end{equation}
where $q^2=-2q\cdot p=-q^2\!\!_\perp$ and $M=m_\ell$ is the mass 
of
the composite system.  The Dirac and Pauli form factors can be
identified\cite{BD,r13} from the spin-conserving and spin-flip current
matrix elements $(J^+=J^0+J^3)$:
\begin{eqnarray}
{\cal M}^+_{\uparrow\uparrow} &=&
\left\langle p+q,\uparrow\left|\frac{J^+(0)}{p^+}\right|p,\uparrow
\right\rangle = 2F_1(q^2) \ ,
\label{eq7}\\[5pt]
{\cal M}^+_{\uparrow\downarrow} &=&
\left\langle p+q,\uparrow\left|\frac{J^+(0)}{p^+}\right|p,\uparrow
\right\rangle = - 2(q_1-iq_2) \ \frac{F_2(q^2)}{2M} \ ,
\label{eq8}
\end{eqnarray}
where $\uparrow$ corresponds to positive spin projection $S_z=+
\frac{1}{2}$ along the $\widehat z$ axis.

Each Fock-state wave function $\left|n\right\rangle$ of the incident
lepton is represented by the functions 
$\psi^{(n)}_{p,S_z}(x_i,\vec
k\!\!_{\perp i},S_i)$, where
\[ k^\mu \equiv (k^+,k^-,\vec k\!\!_\perp) = \left( xp^+,\,
\frac{k^2_\perp+m^2}{xp^+},\vec k\!\!_\perp\right)
\]
specifies the light-cone momentum coordinates of each constituent
$i=1,\ldots,n$, and $S_i$ specifies its spin projection $S^i_z$.
Momentum observation on the light cone requires
\[ \sum^n_{i=1} k_{\perp i} = 0 \ , \qquad
\sum^n_{i=1} x_i= 1 \ ,
\]
and thus $0<x_i<1$.  The amplitude to find $n$ (on-mass-shell)
constituents in the lepton is then $\psi^{(n)}$ multiplied by the
spinor factors $u_{S_i}(k_i)/(k^+_i)^{-1/2}$ or
$v_{S_i}(k_i)/(k^+_i)^{1/2}$ for each constituent fermion or
anti-fermion.   The Fock state is off the ``energy shell'':
\[ \left(p^--\sum^n_{i=1}k^-_i\right) p^+ = \sum^n_{i=1}
\left(\frac{\vec k^2\!\!_{\perp i}+m^2_i}{x_i}\right) \ .
\]
The quantity $(\vec k^2\!\!_{\perp i}+m^2_i)/x_i$ is the
relativistic analog of the kinetic energy $\vec p^2_i/2m_i$ in
the Schr\"odinger formalism.

The wave function for the lepton directed along the final direction
$p+q$ in the current matrix element is then
\[ \psi^{(n)}_{p+q,S^\prime_z} (x_i,\vec k^\prime\!\!_{\perp i},
S^\prime_i) \ ,
\]
where\cite{dy70}
\[ \vec k^\prime\!\!_{\perp j}= \vec k\!\!_{\perp j}
+(1-x_j)\vec q\!\!_\perp
\]
for the struck constituent and
\[ \vec k^\prime\!\!_{\perp i} = \vec k\!\!_{\perp i} -
x_i\vec q\!\!_\perp
\]
for each spectator $(i\ne j)$.  The $\vec k{}^\prime\!\!_\perp$
are transverse to the $p+q$ direction with
\[ \sum^n_{i=1} \vec k^\prime\!\!_{\perp i} = 0 \ .
\]

The interaction of the current $J^+(0)$ conserves the spin
projection of the struck constituent fermion $(\bar
u_s,\gamma^+u_s)/k_+=2\delta_{ss^\prime}$.  Thus from Eqs.
(\ref{eq7}) and (\ref{eq8})
\begin{equation}
F_1(q^2) = \frac{1}{2} {\cal M}^+_{\uparrow\uparrow} =
\sum_j e_j \int [dx]\,\left[d^2\vec k\!\!_\perp\right]\,
\psi^{*(n)}_{p+q,\uparrow}\left(x,\vec
k^\prime\!\!_\perp,S\right)\,
\psi^{(n)}_{p,\uparrow}\left(x,\vec k\!\!_\perp,S\right) \ ,
\label{eq9} 
\end{equation}
and
\begin{eqnarray}
\lefteqn{-\left(\frac{q_1-iq_2}{2M}\right) F_1(q^2)  = 
\frac{1}{2} {\cal M}^+_{\uparrow\downarrow}\nonumber}
\hspace{30pt} \\[5pt] &=&
\sum_j e_j \int [dx]\,\left[d^2\vec k\!\!_\perp\right]\,
\psi^{*(n)}_{p+q,\uparrow}\left(x,\vec
k^\prime\!\!_\perp,S\right)\,
\psi^{(n)}_{p,\uparrow}\left(x,\vec k\!\!_\perp,S\right) \ ,
\label{eq10}
\end{eqnarray}
where $e_j$ is the fractional charge of each constituent.  [A
summation of all possible Fock states $(n)$ and spins $(S)$ is
assumed.]  The phase-space integration is
\begin{equation}
[dx] \equiv \delta \left(1-\sum x_i\right) \prod^n_{i=1} dx_i \ ,
\label{eq11}
\end{equation}
and
\begin{equation}
\left[ d^2k_\perp\right] \equiv 16\pi^3\delta^{(2)}
\left(\sum k_{\perp i}\right) \prod^n_{i=1}
\frac{d^2k_\perp}{16\pi^3} \ .
\label{eq12}
\end{equation}
Equation (\ref{eq9}) evaluated at $q^2=0$ with $F_1(0)=1$ is
equivalent to wavefunction normalization.  The anomalous moment
$a=F_2(0)/F_1(0)$ can be determined from the coefficient linear in
$q_1-iq_2$ from the coefficient linear in $q_1-iq_2$ from
$\psi^*_{p+q}$ in Eq. (\ref{eq10}).  In fact, 
\begin{equation}
\frac{\partial}{\partial\vec q\!\!_\perp}\, \psi^*_{p+q}
\equiv - \sum_{i\ne j}x_i \frac{\partial}{\partial\vec
k\!\!_{\perp i}}\, \psi^*_{p+q}
\label{eq13}
\end{equation}
(summed over spectators), we can, after integration by parts, write
explicitly
\begin{equation}
\frac{a}{M} = - \sum_je_j\int [dx]\int \left[d^2k_\perp\right]
\sum_{i\ne j} \psi^*_{p \uparrow}x_i \left(\frac{\partial}{\partial
k_{1i}}+i\frac{\partial}{\partial k_{2i}}\right) \psi_{p\downarrow} \
. 
\label{eq14}
\end{equation}
The wave function normalization is
\begin{equation}
\int [dx]\int \left[d^2k_\perp\right]\, \psi^*_{p\uparrow}\,
\psi_{p\uparrow} = \int [dx]\int d^2k_\perp\psi^*_{p\downarrow}\,
\psi_{p\downarrow} = 1 \ .
\label{eq15}
\end{equation}
A sum over all contributing Fock states is assumed in Eqs.
(\ref{eq14}) and (\ref{eq15}).
We thus can express the anomalous moment in terms of a local 
matrix
element at zero momentum transfer.  It should be emphasized that 
Eq.
(\ref{eq14}) is exact; it is valid for the anomalous element of any
spin-$\frac{1}{2}$ system.

In the case of the electron's anomalous moment to
order $\alpha$ in QED,\cite{BD,r18} the contributing intermediate Fock
states  are the electron-photon states with spins
$\left| -\frac{1}{2},1\right\rangle$ and
$\left|\frac{1}{2},-1\right\rangle$:
\begin{equation}
\psi_{p\downarrow} = \frac{e/\sqrt x}
{M^2-\frac{k^2_\perp+\lambda^2}{x}-\frac{k^2_\perp+\widehat 
m^2}{1-x}}
\times\left\{\begin{array}{l}
\sqrt 2\  \frac{(k_1-ik_2)}{x}\
\left(\left|-\frac{1}{2}\right\rangle\rightarrow\left|-\frac{1}{2},1
\right \rangle\right) \\[5pt]
\sqrt 2\  \frac{M(1-x)-\widehat m}{1-x}\
\left(\left|-\frac{1}{2}\right\rangle\rightarrow\left|\frac{1}{2},-1
\right\rangle\right)
\end{array} \right.
\label{eq16}
\end{equation}
and
\begin{equation}
\psi^*_{p\uparrow} = \frac{e/\sqrt x}
{M^2-\frac{k^2_\perp+\lambda^2}{x}-\frac{k^2_\perp+
\widehat m^2}{1-x}} \times\left\{
\begin{array}{l}
-\sqrt 2\  \frac{M(1-x)-\widehat m}{1-x}\
\left(\left|-\frac{1}{2},1\right\rangle\rightarrow\left|
\frac{1}{2}\right\rangle\right)\\[5pt]
-\sqrt 2\  \frac{(k_1-ik_2)}{x}\
\left(\left|\frac{1}{2},-1\right\rangle
\rightarrow \left|\frac{1}{2}\right\rangle\right) \ .
\end{array} \right.
\label{eq17}
\end{equation}
The quantities to the left of the curly bracket in Eqs. (\ref{eq16})
and (\ref{eq17}) are the matrix elements of
\[ \frac{\bar u}{(p^+-k^+)^{1/2}}\, \gamma\cdot\epsilon^*\,
\frac{u}{(p^+)^{1/2}}\quad \hbox{and}\quad
\frac{\bar u}{(p^+)^{1/2}}\, \gamma\cdot\epsilon\,
\frac{u}{(p^+-k^+)^{1/2}} \ ,
\]
respectively, where $\widehat \epsilon =
\widehat\epsilon_{\uparrow(\downarrow)}=\pm (1/\sqrt 
2)(\widehat
x\pm i\widehat y)$, $\epsilon\cdot k=0$, $\epsilon^+=0$ in the
light-cone gauge for vector spin projection $S_z=\pm 1$.\cite{r12,r13}
 For the sake of generality, we let the intermediate
lepton and vector boson have mass $\widehat m$ and $\lambda$,
respectively.

Substituting (\ref{eq16}) and (\ref{eq17}) into Eq. (\ref{eq14}),
one finds that only the $\left|-\frac{1}{2},1\right\rangle$
intermediate state actually contributes to $a$, since terms which
involve differentiation of the denominator of $\psi_{p\downarrow}$
cancel.  We thus have\cite{BD}
\begin{equation}
a = 4M\, e^2\int \frac{d^2k_\perp}{16\pi^3}\int^1_0dx\
\frac{\left[\widehat
m-(1-x)M\right]/x(1-x)}{\left[M^2-(k^2_\perp+\widehat
m^2)/(1-x)-(k^2_\perp+\lambda^2)/x\right]^2} \ ,
\label{eq18}
\end{equation}
or
\begin{equation}
a = \frac{\alpha}{\pi}\int^1_0dx\
\frac{M\left[\widehat m-M(1-x)\right]x(1-x)}{\widehat
m^2x+\lambda^2(1-x)-M^2x(1-x)} \ ,
\label{eq19}
\end{equation}
which, in the case of QED $(\widehat m=M, \lambda=0)$ gives the
Schwinger results $a=\alpha/2\pi$.

The general result (\ref{eq14}) can also be written in matrix form:
\begin{equation}
\frac{a}{2M} = - \sum_j e_j \int [dx] \left[d^2k_\perp\right]\,
\psi^+\longvec S\!\!_\perp\cdot\longvec L\!\!_\perp\psi \ ,
\label{eq20}
\end{equation}
where $S$ is the spin operator for the total system and $\longvec
L\!\!_\perp$ is the generator of ``Galiean'' transverse boosts
\cite{r12,r13} on the light cone, \em \ie, \rm $\longvec
S\!\!_\perp\cdot \longvec L\!\!_\perp = (S_+L_-+S_-L_+)/2$ where
$S_\pm = (S_1\pm iS_2)$ is the spin-ladder operator and
\begin{equation}
L_\pm = \sum_{i\ne j} x_i \left(\frac{\partial}{\partial k_{\perp
i}} \mp i\, \frac{\partial}{\partial k_{2i}}\right)
\label{eq21}
\end{equation}
(summed over spectators) in the analog of the angular momentum
operator $\longvec p\times \longvec r$.  Equation (\ref{eq14}) can
also be written simply as an expectation value in impact space.

The results given in Eqs. (\ref{eq9}), (\ref{eq10}), and
(\ref{eq14}) are also valid for calculating the anomalous
moments and form factors of hadrons in quantum chromodynamics
directly from the quark and gluon wave functions $\psi(\longvec
k\!\!_\perp, x, S)$.  These wave functions can also be used to
construct the structure functions and distribution amplitudes which
control large momentum transfer inclusive and exclusive processes.
\cite {r13,r19}  The charge radius of a composite system can also
be written in the form of a local, forward matrix element:
\begin{equation}
\frac{\partial F_1(q^2)}{\partial q^2}\Bigg|_{q^2=0} = - \sum _j
e_j \int [dx]\, \left[d^2k_\perp\right]\, \psi^*_{p,\uparrow}
\left(\sum_{i\ne j} x_i\, \frac{\partial}{\partial\longvec
k\!\!_{\perp i}}\right)^2 \psi_{p,\uparrow} \ .
\label{eq22}
\end{equation}

\section{Magnetic and Electroweak Moments of Nucleons in the Light-Cone Formalism}

The use of covariant kinematics leads to a number of striking
conclusions for the electromagnetic and weak moments of nucleons 
and
nuclei. For example, magnetic moments cannot be written as the 
naive
sum $\overrightarrow\mu = \sum\overrightarrow\mu_i$ of the 
magnetic
moments of the constituents, except in the nonrelativistic limit
where the radius of the bound state is much larger than its Compton
scale: $R_A M_A\gg 1$. The deuteron quadrupole moment is in 
general
nonzero even if the nucleon-nucleon bound state has no $D$-wave
component.\cite{bh83}   
The breakdown of simple additivity for moments and the 
contradictions with the traditional nonrelativistic formalism, 
even for weak binding, is due to the fact that the so-called
``static'' moments must be computed as transitions between states of
different momentum $p^\mu$ and $p^\mu + q^\mu$, with $q^\mu
\rightarrow 0$. Thus one must construct current matrix elements
between boosted states. The Wigner boost generates nontrivial
corrections to the current interactions of bound systems.
\cite{bp69}   Remarkably, in the case of the deuteron, both the
quadrupole and magnetic moments become equal to that of the 
Standard
Model in the limit $M_d R_d\rightarrow 0.$ In this limit, the three
form factors of the deuteron have the same ratios as do those of the
$W$ boson in the Standard Model.\cite{bh83}

One can also use light-cone methods to show that the proton's
magnetic moment $\mu_p$ and its axial-vector coupling $g_A$ have 
a
relationship independent of the specific form of the light-cone
wavefunction.\cite{bs94}  At the physical value of the proton
radius computed from the slope of the Dirac form factor, $R_1=0.76$
fm, one obtains the experimental values for both $\mu_p$ and 
$g_A$;
the helicity carried by the valence $u$ and $d$ quarks are each
reduced by a factor $\simeq 0.75$ relative to their nonrelativistic
values. At infinitely small radius $R_p M_p\rightarrow 0$, $\mu_p$
becomes equal to the Dirac moment, as demanded by the
Drell-Hearn-Gerasimov sum rule.\cite{gerasimov65,dh66}  Another
surprising fact is that as $R_1 \rightarrow 0$ the constituent quark
helicities become completely disoriented and $g_A \rightarrow 0$.

One can understand the origins of the above universal features even
in an effective three-quark light-cone Fock description of the
nucleon. In such a model, one assumes that additional degrees of
freedom (including zero modes) can be parameterized through an
effective potential.\cite{r13}  After truncation, one could in
principle obtain the mass $M$ and light-cone wavefunction of the
three-quark bound-states by solving the Hamiltonian eigenvalue
problem. It is reasonable to assume that adding more quark and
gluonic excitations will only refine this initial approximation.\cite{phw90}
 In such a theory the constituent quarks will also
acquire effective masses and form factors.

Since we do not have an explicit representation for the effective
potential in the light-cone Hamiltonian $P^-_{\rm eff}$ for three
quarks, we shall proceed by making an Ansatz for the momentum-space
structure of the wavefunction $\Psi$.  Even without explicit
solutions of the Hamiltonian eigenvalue problem, one knows that the
helicity and flavor structure of the baryon eigenfunctions will
reflect the assumed global SU(6) symmetry and Lorentz invariance of
the theory.  As we will show below, for a given size of the proton
the predictions and interrelations between observables at $Q^2=0,$
such as the proton magnetic moment $\mu_p$ and its axial coupling
$g_A,$ turn out to be essentially independent of the shape of the
wavefunction.\cite{bs94}

The light-cone model given by Ma \cite{Ma91}  and by Schlumpf
\cite{schlumpf93} provides a framework for representing the 
general
structure of the effective three-quark wavefunctions for baryons.
The wavefunction $\Psi$ is constructed as the product of a 
momentum
wavefunction, which is spherically symmetric and invariant under
permutations, and a spin-isospin wave function, which is uniquely
determined by SU(6)-symmetry requirements.  A Wigner-Melosh 
rotation
\cite{wigner39,melosh74} is applied to the spinors, so that the
wavefunction of the proton is an eigenfunction of $J$ and $J_z$ in
its rest frame.\cite{strik,coester82,ls78}  To represent the range
of uncertainty in the possible form of the momentum wavefunction,
one can choose two simple functions of the invariant mass ${\cal M}$
of the quarks:
\begin{eqnarray} 
\psi_{\rm H.O.}({\cal M}^2) &=& N_{\rm H.O.}\exp(-{\cal
M}^2/2\beta^2),\\ \psi_{\rm Power}({\cal M}^2) &=& N_{\rm Power}
(1+{\cal M}^2/\beta^2)^{-p}\; ,
\end{eqnarray} 
where $\beta$ sets the characteristic internal momentum scale.
Perturbative QCD predicts a nominal power-law fall off at large
$k_\perp$ corresponding to $p=3.5$.\cite{r13}  The Melosh rotation
insures that the nucleon has $j=\frac{1}{2}$ in its rest system.  It
has the matrix representation \cite{melosh74}
\begin{equation}
R_M(x_i,k_{\perp i},m)=\frac{m+x_i {\cal M}-i\overrightarrow
\sigma\cdot(\vec n\times\vec k_i)}{\sqrt{(m+x_i {\cal M})^2+
k_{\perp i}^2} } 
\end{equation} 
with $\vec n=(0,0,1)$, and it becomes the unit matrix if the quarks
are collinear, $R_M(x_i,0,m)=1.$ Thus the internal transverse
momentum dependence of the light-cone wavefunctions also affects 
its
helicity structure.\cite{bp69}

As we showed in Section 6, the Dirac and Pauli form factors $F_1(Q^2)$ 
and $F_2(Q^2)$ of the
nucleons are given by the spin-conserving and the spin-flip matrix
elements of the vector current $J^+_V$ (at $Q^2=-q^2$) \cite{BD}
\begin{eqnarray}
F_1(Q^2) &=& \langle p+q,\uparrow | J^+_V |
p,\uparrow \rangle , \\
(Q_1-i Q_2) F_2(Q^2) &=& -2M\langle
p+q,\uparrow | J^+_V | p, \downarrow \rangle \; .
\end{eqnarray}
We then can calculate the anomalous magnetic moment
$a=\lim_{Q^2\rightarrow 0} F_2(Q^2)$.\footnote{The total proton
magnetic moment is $\mu_p = \frac{e}{2M}(1+a_p).$} The same
parameters as given by Schlumpf \cite{schlumpf93} are chosen, 
namely
$m=0.263$ GeV (0.26 GeV) for the up (down) quark masses,
$\beta=0.607$ GeV (0.55 GeV) for $\psi_{\rm Power}$ ($\psi_{\rm
H.O.}$), and $p=3.5$. The quark currents are taken as elementary
currents with Dirac moments $\frac{e_q}{2 m_q}.$ All of the baryon
moments are well-fit if one takes the strange quark mass as 0.38
GeV. With the above values, the proton magnetic moment is 2.81
nuclear magnetons, and the neutron magnetic moment is $-1.66$
nuclear magnetons. (The neutron value can be improved by relaxing
the assumption of isospin symmetry.) The radius of the proton is
0.76 fm, \ie, $M_p R_1=3.63$.

\begin{figure}[htbp]
\begin{center}
{\epsfxsize 5in\epsfbox{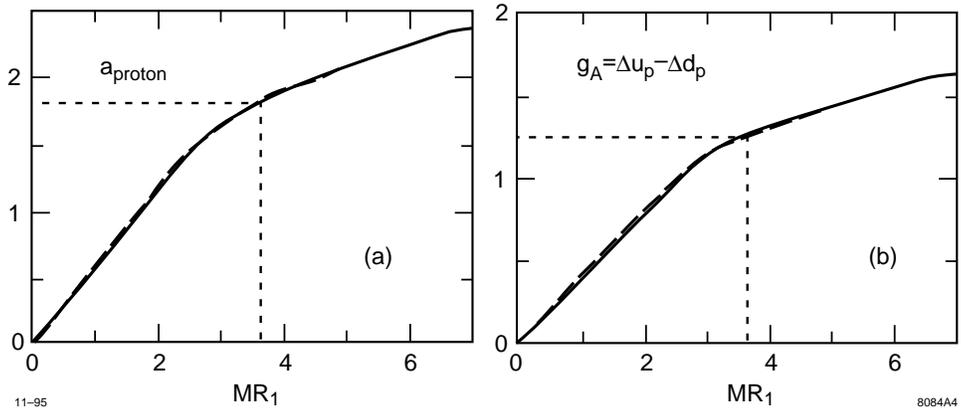}}
\end{center}
\caption[*]{(a).
The anomalous magnetic moment of the proton $a_p=F_2(0)$ as  a
function of its Dirac radius $M_p R_1 $ in Compton units. (b). The
axial vector coupling of the neutron to proton beta-decay as a
function of $M_p R_1.$  In each figure, the broken line is computed
from a wavefunction with power-law fall off  and the solid curve is
computed from a Gaussian wavefunction. The experimental values at
the physical proton Dirac radius are indicated by the dotted line.}
\label{figSchlumpf}
\end{figure}

In Fig.~\ref{figSchlumpf}(a) we show the functional relationship between
the anomalous moment $a_p$ and its Dirac radius predicted by the
three-quark light-cone model. The value of
\begin{equation}
R^2_1 = -6 \frac{dF_1(Q^2)}{dQ^2}\Bigl\vert_{Q^2=0}
\end{equation}
is varied by changing $\beta$ in the light-cone wavefunction while
keeping the quark mass $m$ fixed.  The prediction for the power-law
wavefunction $\psi_{\rm Power}$ is given by the broken line; the
continuous line represents $\psi_{\rm H.O.}$.  Figure~\ref{figSchlumpf}(a)
shows that when one plots the dimensionless observable $a_p$ 
against
the dimensionless observable $M R_1$ the prediction is essentially
independent of the assumed power-law or Gaussian form of the
three-quark light-cone wavefunction.  Different values of $p>2$ also
do not affect the functional dependence of $a_p(M_p R_1)$ shown in
Fig.~\ref{figSchlumpf}(a). In this sense the predictions of the three-quark
light-cone model relating the $Q^2 \rightarrow 0$ observables are
essentially model-independent. The only parameter controlling the
relation between the dimensionless observables in the light-cone
three-quark model is $m/M_p$ which is set to 0.28. For the physical
proton radius $M_p R_1=3.63$ one obtains the empirical value for
$a_p=1.79$ (indicated by the dotted lines in Fig. \ref{figSchlumpf}(a)).

The prediction for the anomalous moment $a$ can be written
analytically as $a=\langle \gamma_V \rangle a^{\rm NR}$, where
$a^{\rm NR}=2M_p/3m$ is the nonrelativistic ($R\rightarrow\infty$)
value and $\gamma_V$ is given as \cite{cc91}
\begin{equation}
\gamma_V(x_i,k_{\perp i},m)= \frac{3m}{{\cal M}}
\left[\frac{(1-x_3){\cal M}(m+x_3 {\cal M})- \vec k_{\perp 3}^2/2}
{(m+x_3 {\cal M})^2+\vec k_{\perp 3}^2}\right]\; .
\end{equation}
The expectation value $\langle \gamma_V \rangle$ is evaluated
as\footnote{Here $[d^3k]\equiv d\vec k_1d\vec k_2d\vec
k_3\delta(\vec k_1+\vec k_2+ \vec k_3)$. The third component of
$\vec k$ is defined as $k_{3i}\equiv\frac{1}{2}(x_i{\cal M}-
\frac{m^2+\vec
k_{\perp i}^2}{x_i {\cal M}})$. This measure differs from the
usual one used \cite{r13} by the Jacobian $\prod
\frac{dk_{3i}}{dx_i}$ which can be absorbed into the wavefunction.}
\begin{equation} 
\langle\gamma_V\rangle = \frac{\int [d^3k] \gamma_V |\psi|^2}{\int
[d^3k] |\psi|^2}\; . 
\end{equation}

Let us now take a closer look at the two limits $R \rightarrow
\infty$ and $R\rightarrow 0$. In the nonrelativistic limit we let
$\beta \rightarrow 0$ and keep the quark mass $m$ and the proton
mass $M_p$ fixed. In this limit the proton radius $R_1 \rightarrow
\infty$ and $a_p \rightarrow 2M_p/3m = 2.38$, since $\langle
\gamma_V \rangle \rightarrow 1$.\footnote{This differs slightly 
from the usual nonrelativistic formula $1+a=\sum_q \frac{e_q}{e}
\frac{M_p}{m_q}$ due to the nonvanishing binding energy which results in 
$M_p\neq 3m_q$.} 
Thus the physical value of the anomalous magnetic
moment at the empirical proton radius $M_p R_1=3.63$ is reduced 
by  25\%\ from its nonrelativistic value due to relativistic recoil and
nonzero $k_\perp$.\footnote{The nonrelativistic value of the neutron
magnetic moment is reduced by 31\%.}

To obtain the ultra-relativistic limit we let $\beta \rightarrow
\infty$ while keeping $m$ fixed.  In this limit the proton becomes
pointlike, $M_p R_1 \rightarrow 0$, and the internal transverse
momenta $k_\perp \rightarrow\infty$. The anomalous magnetic 
momentum
of the proton goes linearly to zero as $a=0.43 M_p R_1$ since
$\langle\gamma_V\rangle\rightarrow 0$.  Indeed, the
Drell-Hearn-Gerasimov sum rule \cite{gerasimov65,dh66} demands 
that
the proton magnetic moment become equal to the Dirac moment at 
small
radius.  For a spin-$\frac{1}{2}$ system
\begin{equation}
a^2=\frac{M^2}{2\pi^2\alpha}\int_{s_{th}}^\infty \frac{ds}{s}\left[
\sigma_P(s)-\sigma_A(s)\right]\; ,
\end{equation}
where $\sigma_{P(A)}$ is the total photo absorption cross section
with parallel (anti-parallel) photon and target spins. If we take
the point-like limit, such that the threshold for inelastic
excitation becomes infinite while the mass of the system is kept
finite, the integral over the photo absorption cross section vanishes
and $a=0$.\cite{BD}  In contrast, the anomalous magnetic moment
of the proton does not vanish in the nonrelativistic quark model as
$R\rightarrow 0$. The nonrelativistic quark model does not reflect
the fact that the magnetic moment of a baryon is derived from 
lepton
scattering at nonzero momentum transfer, \ie, the calculation of a
magnetic moment requires knowledge of the boosted wavefunction.  
The
Melosh transformation is also essential for deriving the DHG sum
rule and low-energy theorems of composite systems.\cite{bp69}

A similar analysis can be performed for the axial-vector coupling
measured in neutron decay. The coupling $g_A$ is given by the
spin-conserving axial current $J_A^+$ matrix element
\begin{equation}
g_A(0) =\langle p,\uparrow | J^+_A | p,\uparrow \rangle\; .
\end{equation}
The value for $g_A$ can be written as $g_A=\langle \gamma_A 
\rangle
g_A^{\rm NR}$, with $g_A^{\rm NR}$ being the nonrelativistic value 
of
$g_A$ and with $\gamma_A$ given by \cite{cc91,ma91}
\begin{equation}
\gamma_A(x_i,k_{\perp i},m)=\frac{(m+x_3 {\cal M})^2-
k_{\perp 3}^2}{(m+x_3 {\cal M})^2+ k_{\perp 3}^2}\; .
\label{gammaa}
\end{equation}
In Fig.~\ref{figSchlumpf}(b) the axial-vector coupling is plotted against
the proton radius $M_p R_1$.  The same parameters and the same 
line
representation as in Fig.~\ref{figSchlumpf}(a) are used.  The functional
dependence of $g_A(M_p R_1)$ is also found to be independent of 
the
assumed wavefunction. At the physical proton radius $M_p 
R_1=3.63$,
one predicts the value $g_A = 1.25$ (indicated by the dotted lines
in Fig.~\ref{figSchlumpf}(b)), since $\langle \gamma_A \rangle =0.75$.  The
measured value is $g_A= 1.2573\pm 0.0028$.\cite{pdg92}   This is a
25\%\ reduction compared to the nonrelativistic SU(6) value
$g_A=5/3,$ which is only valid for a proton with large radius $R_1
\gg 1/M_p.$ The Melosh rotation generated by the internal 
transverse
momentum \cite{ma91} spoils the usual identification of the
$\gamma^+ \gamma_5$ quark current matrix element with the total
rest-frame spin projection $s_z$, thus resulting in a reduction of
$g_A$.

Thus, given the empirical values for the proton's anomalous moment
$a_p$ and radius $M_p R_1,$ its axial-vector coupling is
automatically fixed at the value $g_A=1.25.$ This is an essentially
model-independent prediction of the three-quark structure of the
proton in QCD.  The Melosh rotation of the light-cone wavefunction
is crucial for reducing the value of the axial coupling from its
nonrelativistic value 5/3 to its empirical value. The near equality
of the ratios $g_A/g_A(R_1 \rightarrow \infty)$ and $a_p/a_p(R_1
\rightarrow \infty)$ as a function of the proton radius $R_1$ shows
the wave-function independence of these quantities.  We emphasize
that at small proton radius the light-cone model predicts not only a
vanishing anomalous moment but also $ \lim_{R_1 \rightarrow 0}
g_A(M_p R_1)=0$.  One can understand this physically: in the zero
radius limit the internal transverse momenta become infinite and the
quark helicities become completely disoriented.  This is in
contradiction with chiral models, which suggest that for a zero
radius composite baryon one should obtain the chiral symmetry 
result
$g_A=1$.

The helicity measures $\Delta u$ and $\Delta d$ of the nucleon each
experience the same reduction as does $g_A$ due to the Melosh
effect. Indeed, the quantity $\Delta q$ is defined by the axial
current matrix element
\begin{equation}
\Delta q=\langle p,\uparrow | \bar
q\gamma^+\gamma_5 q | p,\uparrow \rangle\; ,
\end{equation}
and the value for $\Delta q$ can be written analytically as $\Delta
q=\langle \gamma_A \rangle \Delta q^{\rm NR}$, with $\Delta 
q^{\rm
NR}$ being the nonrelativistic or naive value of $\Delta q$ and
$\gamma_A$ given by Eq. (\ref{gammaa}).

The light-cone model also predicts that the quark helicity sum
$\Delta\Sigma=\Delta u+\Delta d$ vanishes as the
proton radius $R_1$ becomes small. Note that  $\Delta\Sigma$ depends on the proton
size, and it should not be identified as the vector sum of the rest-frame
constituent spins. The rest-frame spin sum is not a Lorentz
invariant for a composite system.\cite{ma91}
Empirically, one can
measure $\Delta q$ from the first moment of the leading-twist
polarized structure function $g_1(x,Q).$ In the light-cone and
parton model descriptions, $\Delta q=\int_0^1 dx [q^\uparrow (x) -
q^\downarrow (x)]$, where $q^\uparrow (x)$ and $q^\downarrow 
(x)$
can be interpreted as the probability for finding a quark or
antiquark with longitudinal momentum fraction $x$ and polarization
parallel or anti-parallel to the proton helicity in the proton's
infinite momentum frame.\cite{r13}  [In the infinite momentum
frame there is no distinction between the quark helicity and its
spin projection $s_z.$] Thus $\Delta q$ refers to the difference of
helicities at fixed light-cone time or at infinite momentum; it
cannot be identified with $q(s_z=+\frac{1}{2})-q(s_z=-\frac{1}{2}),$ 
the
spin carried by each quark flavor in the proton rest frame in the
equal-time formalism.

Thus the usual SU(6) values $\Delta u^{\rm NR}=4/3$ and $\Delta
d^{\rm NR}=-1/3$ are only valid predictions for the proton at large
$M R_1.$ At the physical radius the quark helicities are reduced by
the same ratio 0.75 as is $g_A/g_A^{\rm NR}$ due to the Melosh
rotation. Qualitative arguments for such a reduction have been given
elsewhere.\cite{karl92,fritzsch90}  For $M_p R_1 = 3.63,$ the
three-quark model predicts $\Delta u=1,$ $\Delta d=-1/4,$ and
$\Delta\Sigma=\Delta u+\Delta d = 0.75$.  Although the gluon
contribution $\Delta G=0$ in our model, the general sum rule
\cite{jm90}
\begin{equation}
\frac{1}{2}\Delta\Sigma +\Delta G+L_z= \frac{1}{2}
\end{equation}
is still satisfied, since the Melosh transformation effectively
contributes to $L_z$.

Suppose one adds polarized gluons to the three-quark light-cone
model. Then the flavor-singlet quark-loop radiative corrections to
the gluon propagator will give an anomalous contribution $\delta
(\Delta q)=-\frac{\alpha_s}{2\pi}\Delta G$ to each light quark
helicity.\cite{Altarelli}  The predicted value of $g_A = \Delta
u-\Delta d$ is of course unchanged. For illustration we shall choose
$\frac{\alpha_s}{2\pi}\Delta G=0.15$. The gluon-enhanced quark 
model
then gives  values  which agree well with the present
experimental values.

In summary, one sees that relativistic effects are crucial for
understanding the spin structure of nucleons. By plotting
dimensionless observables against dimensionless observables, we
obtain relations that are independent of the momentum-space form 
of
the three-quark light-cone wavefunctions. For example, the value of
$g_A \simeq 1.25$ is correctly predicted from the empirical value of
the proton's anomalous moment. For the physical proton radius $M_p
R_1= 3.63$, the inclusion of the Wigner-Melosh rotation due to the
finite relative transverse momenta of the three quarks results in a
$\sim 25\%$ reduction of the nonrelativistic predictions for the
anomalous magnetic moment, the axial vector coupling, and the 
quark
helicity content of the proton.  At zero radius, the quark
helicities become completely disoriented because of the large
internal momenta, resulting in the vanishing of $g_A$ and the total
quark helicity $\Delta \Sigma.$

\section
{Constructing Hadron Wavefunctions in Light-Cone Quantized QCD }

Our ultimate goal is to actually calculate the light cone wavefunctions of
the hadrons.  In the next two sections, I will discuss possible methods in which one can
obtain constraints and determine important properties  of the wavefunctions, even
in the absence of explicit solutions.  

A remarkable feature of collinear QCD is that although the theory is
effectively one-space and one-time, one still retains the two physical degrees of
freedom from the transversely-polarized gluons.  Thus the spectrum of collinear
QCD contains gluonium states, as well as gluonic quanta in the higher Fock states
of the mesons and baryons eigenstates of the theory.  We have also seen that some
of the features of the structure functions of hadrons in collinear QCD match well
to the phenomenological features of QCD[3+1] such as the
helicity retention of the leading constituents at large $ x \to 1$ in the polarized
structure functions.

Recently Antonuccio, Pinsky and I have investigated the possibility
that one may be able to construct useful models of the light-cone wavefunctions
of QCD[3+1] by extension of the collinear QCD solutions.  We have been
considering two methods:

(1) {\it Minimal Subtraction}.
Let us ignore the complications of spin and write the solution to the
$n$-quark/antiquark light-cone wavefunction of a hadron in the collinear theory in the form
\[ \Psi_n \left[\frac{m^2_i+g^2A^2_\perp}{x_i}\, ,\,
\longvec A_\perp\right]
\]
where the functional dependence of the operator $\Psi_n$ in the field variable
$\longvec A_\perp$ connects Fock states of different gluon number.
The natural generalization of this dependence to the transverse space
dependence is
\[ \Psi_n\rightarrow \Psi_n
\left[\frac{m^2_i + (\longvec k_\perp-g\longvec A_\perp)^2}{x_i}\, ,\,
\left(\longvec k_\perp-g \longvec A_\perp\right)\right]
\]
It is interesting to note that the mechanical light-cone kinetic energy
\[\frac{m^2_i + (\longvec k_\perp-g\longvec A_\perp)^2}{x_i}\, \] is the
essential variable which controls the dynamics of gauge theory.  This is in
agreement with the fact that in laser physics the effective mass of an
electron in an intense laser beam is $m_{\rm effective}^2 = m_e^2+e^2 A^2.$
The laser analog also suggest that a classical approximation to the
gauge field may be useful when the particle number is high.  This could be
appropriate when analyzing the physics of small $x$.

(2) {\it The Light-Cone Lippmann-Schwinger Equation}.
In principle, we can also construct the wavefunctions of QCD(3+1)
 starting with collinear QCD(1+1) solutions by systematic perturbation
theory in $\Delta H$, where $\Delta H$ contains the terms linear and
quadratic in the transverse momenta $\longvec k_{\perp i}$ which are
neglected in the Hamilton $H_0$ of collinear QCD.  We can write the exact
eigensolution of the full Hamiltonian as
\[ \psi_{(3+1)} = \psi_{(1+1)} + \frac{1}{M^2-H + i \epsilon }\,
\Delta H\, \psi_{(1+1)} \ , \]
where
\[\frac{1}{M^2-H + i \epsilon }
 = {\frac{1}{M^2-H_0 + i \epsilon }} +
{\frac{1}{M^2-H+ i \epsilon }}\Delta H{\frac{1}{M^2-H_0 + i \epsilon }} \]
can be represented as the continued iteration of the Lippmann Schwinger
resolvant.
Note that the matrix
$(M^2-H_0)^{-1}$ is known to any desired precision from the DLCQ solution of collinear QCD.

In each of these methods, the resulting wavefunction can be considered
as approximate solution to 3+1 QCD hadron wavefunctions which could be
subsequently improved by variational or other methods.

\section{Determining the Far Off-Shell Behavior of Hadron LC Wavefunctions}

In many cases of physical interest we are specifically interested in the behavior of
the LC wavefunctions in the far-off-shell domain where ${\cal M}^2_n$ exceeds the
global cutoff.  This occurs for large parton momenta, $k^2_\perp \rightarrow
\infty$,
$x_i\rightarrow 1$ and for massive quanti-antiquark fluctuations.  In fact, in
such domains, we can construct the wavefunction perturbatively and obtain
rigorous QCD predictions.

The basic method is as follows.\cite{BrodskyLepage}  Suppose we can solve
\begin{equation} H_{LC}(\mu) \ket{\psi^{(\mu)}} = M^2 \ket{\psi^{(\mu)}}
\end{equation} in the soft-domain with ${\cal M}^2_n < \mu^2$, where $\mu^2$ is
of the order of a few GeV$^2$.  Then
\begin{equation}
\ket{\psi^{(\mu)}} = \sum_n\ket n \VEV{n\,|\,\psi}\, \theta(\mu^2-M^2_n)
\equiv P^{(\mu)}\ket\psi \ .
\end{equation} 
We can also define the complement projection operator $Q^{(\mu)}$
with $P^{(\mu)}+Q^{(\mu)}=1$.  The full solution satisfies $H\ket\psi =
M^2\ket\psi$ or
\begin{eqnarray} PHP\ket\psi + PHQ\ket\psi &=& M^2P\ket\psi \nonumber \\[2ex]
QHP \ket\psi + QHQ\ket\psi &=& M^2Q\ket\psi \ .
\end{eqnarray} 
Therefore
\begin{equation}
\ket\psi = \ket{\psi^{(\mu)}} + \frac{1}{M^2-QHQ}\ QH_\pm \ket{\psi^{(\mu)}}
\end{equation} 
where only high mass $(\M^2_n > \mu^2)$ perturbatively calculable
intermediate states appear.  An example is the behavior
of the valence wavefunction at large internal transverse momentum.  One finds at
large
$\vec k_\perp = \vec q_\perp$
\begin{eqnarray}
\psi(x,q_\perp) &=& \frac{1}{m^2-\frac{\vec q^2_\perp+m^2}{x(1-x)}}
\int^1_0dy\int d^2\ell_\perp\, V(x,q_\perp;y,\ell_\perp)\, \psi^{(\mu)}
(y,\ell_\perp) \nonumber\\[2ex] &\cong& - \frac{x(1-x)}{\vec q^2_\perp}\
\alpha_s(\vec q^2_\perp)
\int^1_0dy\, V_{1g}(x,y)\, \psi (y,q^2_\perp)
\end{eqnarray} where
\begin{equation}
\phi(y,Q^2) = \int^{Q^2} \frac{d^3\ell_\perp}{16\pi^3}\
\psi_{(2)}(y,\ell_\perp)
\end{equation} is the meson distribution amplitude, which takes the role of the
wavefunction ``at-the-origin'' in analogous nonrelativistic calculations.  The
simple $\alpha_s(q^2_\perp)/\vec q^2_\perp$ fall-off of the valence wavefunction
is the key input to the derivation of dimensional counting rules for form
factors and other exclusive processes in QCD.\cite{BrodskyFarrar}  One can also derive the evolution equation for 
\begin{equation}
\frac{\partial\phi(y,Q^2)}{\partial\log Q^2}
\end{equation} from this result and the simple properties of the one-gluon
exchange kernel.

\section{Structure Functions at the End-Point}

The $x\rightarrow 1$ behavior of quark and gluon wavefunctions is controlled by
far-off-shell configuration and thus can be analyzed perturbatively.  The
dominant contributions come from the lowest Fock states which contain the
partons.  For example the quark distribution in the nucleon can be computed from
two iterations of the gluon exchange kernel which transfers the light-cone
momentum to the struck quark from the two spectators which are required to stop. 
The result is a nominal power-law fall-off at $x\rightarrow 1$: $q_\uparrow(x)
\sim (1-x)^3$ for quarks with helicity aligned with the proton and $q_\downarrow
\sim (1-x)^5$ for quarks anti-aligned.  Similarly the $\ket{qqqg}$ Fock state
yields
$g_\uparrow(x)\sim (1-x)^4$ for gluons with helicity aligned and
$g_\downarrow\sim (1-x)^6$ for gluons helicity anti-aligned.  Thus the partons
with $x\rightarrow 1$ tend to have the same sign helicity as the bound state.\cite{BrodskyLepage}

\begin{figure}[htbp]
\begin{center}
\leavevmode
{\epsfxsize=3in \epsfbox{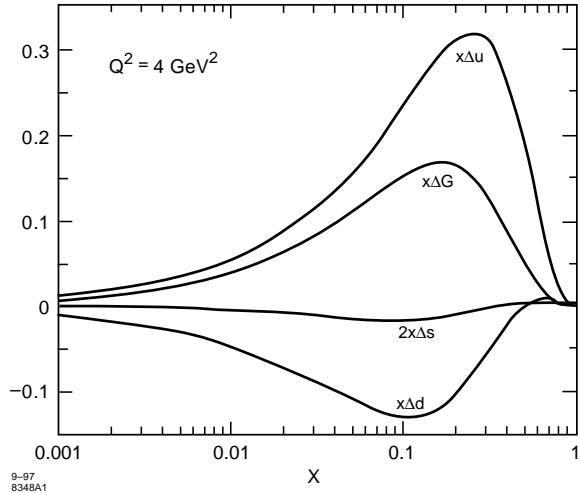}}
\end{center}
\caption[*]{Model form for the polarized quark and gluon distributions of the 
proton satisfying empirical constraints and the perturbative QCD and Regge input conditions.\cite {BBS} 
From Leader \etal\cite {Leader}}
\label{fig:Leader}
\end{figure}

In the absence of full solutions to the light-cone wavefunctions, one can
construct a simple phenomenology of the polarized and anti-polarized structure
functions by imposing a smooth connection between the perturbative QCD constraints at
$x\rightarrow 1$, and Regge behavior at $x\rightarrow 0$, and the momentum and
Bjorken sum rules.  A complete discussion can be found in the literature.\cite{BBS}
Recently Leader \etal\ have shown that this parametrization agrees remarkably
well with the available data from SLAC and CERN.\cite {Leader} 
Figure \ref{fig:Leader} shows the
nominal form of the helicity distributions $\Delta q(x) =
q_\uparrow(x)-q_\downarrow(x)$, for the valance quarks and gluons in the proton.

\vspace{.5cm}

\section{Intrinsic Hardness}

The light-cone wavefunctions contain high fluctuations of arbitrary mass; \ie\
nonzero probabilities for massive pairs, massive sea quarks, etc.  These
fluctuations are of two types:  extrinsic $qg$, $gg$ or $q\bar q$ high mass
pairs which are associated with the substructure of the constituents and are
contained in ordinary DGLAP evolution; and intrinsic functions which are due to
the physics of the bound state wavefunction itself.  For example intrinsic sea
quark pairs $Q\bar Q$ arise from diagrams which are interconnected to the
valence quarks of the bound state hadron and thus depend on the valence quark
correlations.  It is easy to see that the leading perturbative contribution to
intrinsic pairs falls as
\begin{equation}
\psi \sim \frac{\alpha_s^{2+\Gamma}(\M^2)}{\M^2}
\end{equation} where $\M$ is the pair mass and $T$ is the leading anomalous
dimension associated with the valence wavefunction.  The probability of any
configurations with $\M^2> \M^2_0$ is then
\begin{equation} P(\M^2> \M^2_0) \sim \frac{\alpha_s^{4+2\Gamma}}{\M^2_0} \ ,
\end{equation} which implies a remarkably slow fall-off for large off-shell
fluctuations.  The result is universal  for any intrinsic parton pair
\begin{equation}
\frac{\M^2+P^2_\perp}{x_1+x_2} = \frac{m^2_1+\vec k^2_{\perp i}}{x_1} +
\frac{m^2_2+\vec k^2_{\perp 2}}{x_2}
\end{equation} with
\begin{equation}
\vec k_{\perp 1} = x_1\vec P_\perp + \vec k_\perp \qquad
\vec k_{\perp 2} = x_2 \vec P_\perp - \vec k_\perp \ ; 
\end{equation}
\ie\ for large relative $k_\perp$ and/or large quark mass.  Hoyer and I
\cite{HB}\ call this ``intrinsic hardness.''  In the case of intrinsic charm or
bottom pairs in the nucleon, the LC wavefunction is maximized when $\M^2_n$ is
minimized; \ie\ for
\begin{equation} x_{\perp i} = \frac{m_{\perp i}}{\Sigma m_{\perp i}}
\end{equation} where $m_\perp = \sqrt{m^2+\vec k^2_\perp}$.  Thus the maximal
intrinsic charm-bottom configurations occur at equal rapidity; \ie\ where the
heavy partons have highest momentum fractions. This is in contrast to the usual
extrinsic sea quark which  are subconstituents of the gluons and have low $x$. 
The extrinsic quarks evolve rapidly with a probability increasing as power of
$\log\, Q^2/\M^2$.

It is thus important to distinguish two types of quark and gluon
contributions to the nucleon sea measured in deep inelastic
lepton-nucleon scattering: ``extrinsic" and ``intrinsic".
\cite{intcxx}  The extrinsic sea quarks and gluons are created as
part of the lepton-scattering interaction and thus exist over a very
short time $\Delta \tau \sim 1/Q$. These factorizable contributions
can be systematically derived from the QCD hard bremsstrahlung and
pair-production (gluon-splitting) subprocesses characteristic of
leading twist perturbative QCD evolution. In contrast, the intrinsic
sea quarks and gluons are multi-connected to the valence quarks and
exist over a relatively long lifetime within the nucleon bound
state. Thus the intrinsic $q \bar q$ pairs can arrange themselves
together with the valence quarks of the target nucleon into the most
energetically-favored meson-baryon fluctuations.

Another interesting distinction between extrinsic and intrinsic sea quarks is
that due to nonperturbative effects, the intrinsic contributions are generally
not symmetric for sea quark and anti-quarks.  For example, in the muonium atom
$(\mu^+e^-)$ an intrinsic
$\tau^+\tau^-$ pair would be asymmetric since the $\tau^+$ tends to be attracted
to the electron and the $\tau^-$ tends to be attracted to the opposite-sign
muon. Thus the $\tau^-$ would be expected to have a higher $\VEV x$ than the
$\tau^+$. It is also possible to consider the nucleon wavefunction at low
resolution as a fluctuating system coupling  to intermediate
hadronic Fock states such as non-interacting meson-baryon pairs. 
The
most important fluctuations are most likely to be those closest to
the energy shell and thus have minimal invariant mass.  For 
example,
the coupling of a proton to a virtual $K^+ \Lambda$ pair provides a
specific source of intrinsic strange quarks and antiquarks in the
proton.   Since the $s$ and $\bar s$ quarks appear in different
configurations in the lowest-lying hadronic pair states, their
helicity and momentum distributions are distinct. Ma and I \cite{MB}\ 
have used an intermediate meson-baryon
fluctuation model to model the possible $s$ versus $\bar s$ and $c$ versus $\bar
c$ asymmetries of the intrinsic distributions of the nucleon. 
We utilize a boost-invariant light-cone Fock
state description of the hadron wavefunction which emphasizes
multi-parton configurations  of minimal invariant mass. We find that
such fluctuations predict a striking sea quark/antiquark
asymmetry in the corresponding momentum and helicity 
distributions
in the nucleon structure functions.  In particular, the strange and
anti-strange distributions in the nucleon generally have completely
different momentum and spin characteristics. The helicity
structure of the intrinsic $s\bar s$ is strongly asymmetric:  the $s$ quark from
a $\Lambda(uds)\, K(u\bar s)$ is aligned with the $\Lambda$ helicity and
(because of parity) is 100\%\ anti-aligned with the nucleon spin.   On the other
hand, the $\bar s$ from the pseudoscalar kaon is unaligned.  Ma and I have shown that this
picture of quark and antiquark asymmetry in the momentum and
helicity distributions of the nucleon sea quarks has support from a
number of experimental observations, and we have suggested processes to
test and measure this quark and antiquark asymmetry in the 
nucleon sea.

\subsection {Phenomenological Consequences of Intrinsic Charm and Bottom }

Microscopically, the intrinsic heavy-quark Fock component in the
$\pi^-$ wavefunction, $ \vert \overline u d Q \overline Q \rangle$,
is generated by virtual interactions such as $g g \rightarrow Q
\overline Q$ where the gluons couple to two or more projectile
valence quarks. The probability for $Q \overline Q$ fluctuations to
exist in a light hadron thus scales as $\alpha_s^2(m_Q^2)/m_Q^2$
relative to leading-twist production.\cite{vbxx}  This contribution
is therefore higher twist, and power-law suppressed compared to sea
quark contributions generated by gluon splitting.  When the
projectile scatters in the target, the coherence of the Fock
components is broken and its fluctuations can hadronize, forming 
new hadronic systems from the fluctuations.\cite{bhmt92}  For example,
intrinsic $c \overline c$ fluctuations can be liberated provided the
system is probed during the characteristic time $\Delta t = 2p_{\rm
lab}/M^2_{c \overline c}$ that such fluctuations exist. For soft
interactions at momentum scale $\mu$, the intrinsic heavy quark
cross section is suppressed by an additional resolving factor
$\propto \mu^2/m^2_Q$.\cite{chevxx}  The nuclear dependence 
arising from the manifestation of intrinsic charm is expected to be
$\sigma_A\approx \sigma_N A^{2/3}$, characteristic of soft
interactions.

In general, the dominant Fock state configurations are not far off
shell and thus have minimal invariant mass ${\cal M}^2 = \sum_i
m_\perp/ x_i$ where $m_\perp$ is the transverse mass of the
$i^{\rm th}$ particle in the configuration. Intrinsic $Q \overline
Q$ Fock components with minimum invariant mass correspond to
configurations with equal-rapidity constituents. Thus, unlike sea
quarks generated from a single parton, intrinsic heavy quarks tend
to carry a larger fraction of the parent momentum than do the light
quarks.\cite{intcxx}  In fact, if the intrinsic $Q \overline Q$
pair coalesces into a quarkonium state, the momentum of the two
heavy quarks is combined so that the quarkonium state will carry a
significant fraction of the projectile momentum.

There is substantial evidence for the existence of intrinsic $c
\overline c$ fluctuations in the wavefunctions of light hadrons. For
example, the charm structure function of the proton measured by 
EMC
is significantly larger than that predicted by photon-gluon fusion
at large $x_{Bj}$.\cite{emcicxx}  Leading charm production in $\pi
N$ and hyperon-$N$ collisions also requires a charm source beyond
leading twist.\cite{vbxx,ssnxx}  The NA3 experiment has also shown
that the single $J/\psi$ cross section at large $x_F$ is greater
than expected from $gg$ and $q \overline q$ production.\cite{badxx}
The nuclear dependence of this forward component is
diffractive-like, as expected from the BHMT mechanism.  In addition,
intrinsic charm may account for the anomalous longitudinal
polarization of the $J/\psi$ at large $x_F$ seen in $\pi N
\rightarrow J/\psi X$ interactions.\cite{vanxx} Further theoretical
work is needed to establish that the data on direct $J/\psi$ and
$\chi_1$ production can be described using the  higher-twist
intrinsic charm mechanism.\cite{bhmt92}

A recent analysis by Harris, Smith and Vogt \cite{HarrisSmithVogt} of the
excessively large charm structure function of the proton at large
$x$ as measured by the EMC collaboration at CERN yields an estimate
that the probability $P_{c \bar c}$ that the proton contains
intrinsic charm Fock states is of the order of $0.6\% \pm 0. 3 \%.$ 
In the case of intrinsic bottom, perturbative QCD scaling predicts
\begin{equation}
P_{ b \bar b}=P_{c \bar c} \frac{m^2_\psi}{m^2_\Upsilon}
\frac{\alpha^4_s(m_b)}{\alpha^4_s(m_c)}\; ,
\end{equation}
more than an order of magnitude smaller.  We can speculate that if super-partners
of the quarks or gluons exist they must also appear in higher Fock states
of the proton, such as $\vert uud ~{\rm gluino}~ {\rm gluino}
\rangle$. At sufficiently high energies, the diffractive excitation
of the proton will produce these intrinsic quarks and gluinos in the
proton fragmentation region.  Such supersymmetric particles can 
bind
with the valence quarks to produce highly unusual color-singlet
hybrid supersymmetric states such as $\vert uud ~{\rm
gluino}\rangle$ at high $x_F.$ The probability that the proton
contains intrinsic gluinos or squarks scales with the appropriate
color factor and scales inversely with the heavy particle mass squared
relative to the intrinsic charm and bottom probabilities.  This
probability is directly reflected in the production rate when the
hadron is probed at a hard scale $Q$ which is large compared to the
virtual mass ${\cal M}$ of the Fock state.  At low virtualities, the
rate is suppressed by an extra resolution factor of $Q^2/{\cal M}^2.$ The
forward proton fragmentation regime is a challenge to instrument at
HERA, but it may be feasible to tag special channels involving
neutral hadrons or muons.  In the case of the gas jet fixed-target
$ep$ collisions  such as at HERMES, the target fragments emerge at
low velocity and large backward angles, and thus may be accessible
to precise measurement.

{\it Double Quarkonium Hadroproduction }
It is quite rare for two charmonium states to be produced in the
same hadronic collision.  However, the NA3 collaboration has
measured a double $J/\psi$ production rate significantly above
background in multi-muon events with $\pi^-$ beams at laboratory
momentum 150 and 280 GeV/c and a 400 GeV/c proton beam.
\cite{badpxx}  The relative double to single rate, $\sigma_{\psi
\psi}/\sigma_\psi$, is $(3 \pm 1) \times 10^{-4}$ for pion-induced
production, where $\sigma_\psi$ is the integrated single $\psi$
production cross section.  A particularly surprising feature of the
NA3 $\pi^-N\rightarrow\psi\psi X$ events is that the laboratory
fraction of the projectile momentum carried by the $\psi \psi$ pair
is always very large, $x_{\psi \psi} \geq 0.6$ at 150 GeV/c and
$x_{\psi \psi} \geq 0.4$ at 280 GeV/c.  In some events, nearly all
of the projectile momentum is carried by the $\psi \psi$ system! In
contrast, perturbative $ g g$ and $q \overline q$ fusion processes
are expected to produce central $\psi \psi$ pairs, centered around
the mean value, $\langle x_{\psi\psi} \rangle \approx$ 0.4--0.5, in
the laboratory.  The predicted
$\psi \psi$ pair distributions from the intrinsic charm model
provide a natural explanation of the strong forward production of
double $J/\psi$ hadroproduction, and thus gives strong
phenomenological support for the presence of intrinsic heavy quark
states in hadrons.

It is clearly important for the double $J/\psi$ measurements to be
repeated with higher statistics and at higher energies. The same
intrinsic Fock states will also lead to the production of
multi-charmed baryons in the proton fragmentation region. The
intrinsic heavy quark model can also be used to predict the features
of heavier quarkonium hadroproduction, such as $\Upsilon 
\Upsilon$,
$\Upsilon \psi$, and $(c\bar b)$ $(\bar cb)$ pairs. Predictions for these events
have been given by Ramona Vogt and myself.

{\it Leading-Particle Effect in Open Charm Production }
According to PQCD factorization, the fragmentation of a heavy quark
jet is independent of the production process. However, there are
strong correlations between the quantum numbers of $D$ mesons 
and
the charge of the incident pion beam in $\pi N \rightarrow D X$
reactions. This effect can be explained as being due to the
coalescence of the produced intrinsic charm quark with co-moving
valence quarks. The same higher-twist recombination effect can also
account for the suppression of $J/\psi$ and $\Upsilon$ production in
nuclear collisions in regions of phase space with high particle
density.\cite{vbxx}

It is of particular interest to examine the fragmentation of the
proton when the electron strikes a light quark and the interacting
Fock component is the $\vert uud c \bar c \rangle$ or $\vert uud b
\bar b \rangle$ state.  These Fock components correspond to
intrinsic charm or intrinsic bottom quarks in the proton
wavefunction.  Since the heavy quarks in the proton bound state 
have
roughly the same rapidity as the proton itself, the intrinsic heavy
quarks will appear in the  proton fragmentation region.  One expects heavy
quarkonium  and
also heavy hadrons to be formed from the coalescence of the heavy
quark with the valence $u$ and $d$ quarks, since they have nearly
the same rapidity.  

\section {Intrinsic Charm and the $J/\psi \to \rho \pi$ problem}
One of the most dramatic problems confronting the standard picture of quarkonium
decays is the $J/\psi \to \rho\pi$ puzzle.\cite{puzzle} This decay occurs with a
branching ratio of $(1.28\pm 0.10)\%$,\cite{RPP96} and it is the largest
two-body hadronic branching ratio of the $J/\psi$. The $J/\psi$ is assumed to be
a $\cbar c$ bound state pair in the $\Psi(1S)$ state. One then expects the
$\psi^\prime = \Psi(2S)$ to decay to $\rho\pi$ with a comparable branching ratio,
scaled by a factor $\sim 0.15$, due to the ratio of the $\Psi(2S)$ to
$\Psi(1S)$ wavefunctions squared at the origin. In fact, $B(\psi^\prime
\to \rho\pi) < 3.6\times 10^{-5}$ ,\cite{BES96} more than a factor of 50
below the expected rate. Most of the branching ratios for exclusive
hadronic channels allowed in both $J/\psi$ and $\psi^\prime$ decays
indeed scale with their lepton pair branching ratios, as would be
expected from decay amplitudes controlled by the quarkonium wavefunction
near the origin,\cite{RPP96,BES96}
\begin{equation}
{B(\psi^\prime \to h)\over B(J/\psi \to h)}\simeq 
{B(\psi^\prime \to e^+e^-)\over B(J/\psi \to e^+e^-)} 
=0.147\pm0.023 \quad
\label{Bscaling}
\end{equation}
where $h$ denotes a given hadronic channel.
The $J/\psi\to \rho\pi$ and $J/\psi \to K K^*$ decays also conflict dramatically
with perturbative QCD hadron helicity conservation: all such pseudoscalar/vector two-body
hadronic final states are forbidden at leading twist if helicity is conserved at
each vertex.\cite{BrodskyLepage,BLT}

Marek Karliner and I \cite {Karliner} have recently shown that 
such anomalously large decay rates
for the 
$J/\psi$ and their suppression for
$\psi^\prime(2S)$ follow naturally from the existence of intrinsic charm
$\ket{\qbar q \, \cbar c}$ Fock components of the light vector mesons.  
For example, consider the
light-cone Fock representation of the $\rho$: $\rho^+ = \Psi^\rho_{u \dbar}
\,\ket{u \dbar} \,+\, \Psi^\rho_{u\dbar\,\cbar c}\,\ket{u \dbar\,\cbar
c}\,+\,\cdots\ \ .$ The
$\Psi^\rho_{u \dbar \cbar c} $ wavefunction will be maximized at minimal
invariant mass; \ie\ at equal rapidity for the constituents and in the spin
configuration where the $u \dbar$ are in a pseudoscalar state, thus minimizing
the QCD spin-spin interaction. The $\cbar c$ in the $\ket{u \dbar\,\cbar c}$
Fock state carries the spin projection of the $\rho.$ We also expect the
wavefunction of the $\cbar c$ quarks to be in an $S$-wave configuration with no
nodes in its radial dependence, in order to minimize the kinetic energy of the
charm quarks and thus also minimize the total invariant mass.

The presence of the $\ket{u \dbar\,\cbar c}$ Fock state in the $\rho$ will allow
the $J/\psi \to \rho\pi$ decay to occur simply through rearrangement of the
incoming and outgoing quark lines; in fact, the $\ket{u \dbar \,\cbar c}$ Fock
state wavefunction has a good overlap with the radial and spin $\ket{\cbar c}$
and $\ket{u \dbar}$ wavefunctions of the $J/\psi$ and pion. 
Moreover, there is no conflict 
with hadron helicity conservation, since the $\cbar c$ pair in the
$\rho$ is in the $1^-$ state. 
On the other hand, the overlap with the
$\psi^\prime$ will be suppressed, since the radial wavefunction of the $n = 2$
quarkonium state is orthogonal to 
the node-less $\cbar c$ in the $\ket{u \dbar \,\cbar c}$ 
state of the $\rho$. This simple argument provides a 
compelling explanation of the absence of $\psi^\prime \to \rho\pi$ and other 
vector pseudoscalar-scalar states.\footnote{The possibility that the radial
configurations of the initial and final states could be playing a role in the
$J/\psi \to \rho\pi$ puzzle was first suggested by S. Pinsky,\cite{Pinsky} who
however had in mind the radial wavefunctions of the light quarks in the $\rho$,
rather than the wavefunction of the $\cbar c$ intrinsic charm components of the
final state mesons.}

\section{Light-Cone Wavefunction Description of the Spin Anomaly in Deep
Inelastic Polarized Structure Functions}

One of the most interesting distinguishing characteristics between extrinsic and
intrinsic heavy quarks is their contributions to the Ellis-Jaffe sum rule
$\int^1_0 dx\, g_1(x,Q)$ for polarized deep inelastic scattering cross
sections.  The extrinsic contributions to structure functions can be identified
with photon-gluon fusion processes since they derive from $Q\bar Q$ constituents
of the gluon. However, one obtains zero contribution to the Ellis-Jaffe sum rule
from
$\gamma^*g\rightarrow q\bar q$ at tree level if the gluon is on-shell $k^2=0$. 
This follows from the DHG sum rule:  the tree graph
contribution to \cite{bs}
\begin{equation}\int^\infty_{\nu_{th}} \frac{d{\nu}}{\nu}\, \Delta\sigma(ab
\rightarrow cd)
\end{equation} vanishes for any two-to-two polarized cross sections if $a$ is an
on-shell gauge particle.  Thus the anomaly contribution
$-\frac{\alpha_s}{2\pi}\, \Delta G$ to the Ellis-Jaffe sum rule arises 
from off-shell gluons with $|k^2| \gsim 4 m^2_Q$.\cite{sjbias}  The final state which
contributes physically to such configurations consists of $Q$ and $\bar Q$ jets
recoiling against the scattered lepton plus a third jet scattering at $p_T > m_Q$,
corresponding to a quark (or gluon) which emitted the off-shell gluon.  The
intrinsic contributions, on the other hand, consist of one high $p_T$ heavy
quark jet recoiling against the lepton.  Also, as noted above, the $\Delta
Q(x)$ and $\Delta\bar Q(x)$ in general are different helicity distributions.\cite{MB}

\section {Direct Measurement of the Light-cone Valence Wavefunction.}

Diffractive multi-jet production in heavy
nuclei provides a novel way to measure the shape of the LC Fock
state wavefunctions. For example, consider the reaction
\cite{Bertsch,MillerFrankfurtStrikman}
\begin{equation}
\pi A \rightarrow {\rm Jet}_1 + {\rm Jet}_2 + A^\prime
\label{eq:h}
\end{equation}
at high energy where the nucleus $A^\prime$ is left intact in its ground
state.  The transverse momenta of the jets have to balance so that
$
\vec k_{\perp i} + \vec k_{\perp 2} = \vec q_\perp < \R^{-1}_A \ ,
$
and the light-cone longitudinal momentum fractions have to add to
$x_1+x_2 \sim 1$ so that $\Delta p_L < \R^{-1}_A$.  The process can
then occur coherently in the nucleus.  Because of color transparency; \ie\
the cancellation of color interactions in a small-size color-singlet
hadron, the valence wavefunction of the pion with small impact
separation, will penetrate the nucleus with minimal interactions,
diffracting into jet pairs.\cite{Bertsch} The $x_1=x$, $x_2=1-x$ dependence of
the di-jet distributions will thus reflect the shape of the pion distribution
amplitude; the $\vec k_{\perp 1}- \vec k_{\perp 2}$
relative transverse momenta of the jets also gives key information on
the underlying shape of the valence pion wavefunction.  The QCD analysis can be
confirmed by the observation that the diffractive nuclear amplitude
extrapolated to $t = 0$ is linear in nuclear number $A$, as predicted by QCD color
transparency.  The integrated diffractive rate should scale as $A^2/\R^2_A \sim
A^{4/3}$. A diffractive experiment of this type is now in progress at Fermilab
using 500 GeV incident pions on nuclear targets.\cite{E791} 

Data from CLEO for the
$\gamma
\gamma^* \rightarrow \pi^0$ transition form factor favor a form for
the pion distribution amplitude close to the asymptotic solution \cite{BrodskyLepage}
$\phi^{asympt}_\pi (x) =
\sqrt 3 f_\pi x(1-x)$ to the perturbative QCD evolution
equation.\cite{Kroll,Rad,BJPR} It will be interesting to see if the diffractive
pion to di-jet experiment also favors the asymptotic form.

It would also be interesting to study diffractive tri-jet production using proton
beams
$ p A \rightarrow {\rm Jet}_1 + {\rm Jet}_2 + {\rm Jet}_3 + A^\prime $ to
determine the fundamental shape of the 3-quark structure of the valence
light-cone wavefunction of the nucleon at small transverse separation. 
Conversely, one can use incident real and virtual photons:
$ \gamma^* A \rightarrow {\rm Jet}_1 + {\rm Jet}_2 + A^\prime $ to
confirm the shape of the calculable light-cone wavefunction for
transversely-polarized and longitudinally-polarized virtual photons.  Such
experiments will open up a remarkable, direct window on the amplitude
structure of hadrons at short distances.

\section{Other Applications of Light-Cone Quantization to Hadron Phenomenology}

The light-cone formalism provides the theoretical framework which
allows for a hadron to exist in various Fock configurations.  For
example, quarkonium states not only have valence $Q \overline Q$
components but they also contain $Q\overline Q g$ and 
$Q \overline Q g g$ states in which the quark pair is in a color-octet
configuration. Similarly, nuclear LC wave functions contain
components in which the quarks are not in color-singlet nucleon
sub-clusters.  In some processes, such as large momentum transfer
exclusive reactions, only the valence color-singlet Fock state of
the scattering hadrons with small inter-quark impact separation
$b_\perp = {\cal O} (1/Q)$ can couple to the hard scattering
amplitude.  In reactions in which large numbers of particles are
produced, the higher Fock components of the LC wavefunction will be
emphasized. The higher particle number Fock states of a hadron
containing heavy quarks can be diffractively excited, leading to
heavy hadron production in the high momentum fragmentation 
region of
the projectile. In some cases the projectile's valence quarks can
coalesce with quarks produced in the collision, producing unusual
leading-particle correlations.  Thus the multi-particle nature of
the LC wavefunction can manifest itself in a number of novel ways. 
For example:

{\it Regge behavior.} The light-cone wavefunctions $\psi_{n/H}$ of a hadron are
not independent of each other, but rather are coupled via the equations of
motion.  Recently Antonuccio, Dalley and I
\cite{ABD} have used the constraint of finite ``mechanical'' kinetic energy to
derive``ladder relations" which interrelate the light-cone wavefunctions of
states differing by 1 or 2 gluons.  We then use these relations to derive the
Regge behavior of both the polarized and unpolarized structure functions at $x
\rightarrow 0$, extending Mueller's derivation of the BFKL hard
QCD pomeron from the properties of heavy quarkonium light-cone wavefunctions at
large
$N_C$ QCD.\cite{Mueller}

{\it Analysis of diffractive vector meson photoproduction.} The
light-cone Fock wavefunction representation of hadronic amplitudes
allows a simple eikonal analysis of diffractive high energy processes, such as
$\gamma^*(Q^2) p \to \rho p$, in terms of the virtual photon and the vector meson
Fock state light-cone wavefunctions convoluted with the $g p \to g p$
near-forward matrix element \cite{BGMFS} 
See Fig. 1h.  
One can easily show that only small
transverse size $b_\perp \sim 1/Q$ of the vector meson wavefunction is involved.
The hadronic interactions are minimal, and thus the $\gamma^*(Q^2) N \to \rho N$
reaction can occur coherently throughout a nuclear target in reactions such as
without absorption or shadowing. The $\gamma^* A \to \phi A$ process thus
provides a natural framework for testing QCD color
transparency.\cite{BM}

{\it Structure functions at large $x_{bj}$.} The behavior of structure functions
where one quark has the entire momentum requires the knowledge of LC wavefunctions
with $x \rightarrow 1$ for the struck quark and $x \rightarrow 0$ for the
spectators.  As mentioned in Section 2, this is a highly off-shell configuration,
and thus one can rigorously derive quark-counting and helicity-retention rules
for the power-law behavior of the polarized and unpolarized quark and gluon
distributions in the $x
\rightarrow 1$ endpoint domain. Evolution
of structure functions is minimal in this domain because the struck quark is highly
virtual as $x\rightarrow 1$; \ie\ the starting point $Q^2_0$ for evolution cannot
be held fixed, but must be larger than a scale of order
$(m^2 + k^2_\perp)/(1-x)$.\cite{BrodskyLepage}

{\it Color Transparency }
QCD predicts that the Fock components of a hadron with a small color
dipole moment can pass through nuclear matter without interactions.\cite{Bertsch,BM}
  Thus in the case of large momentum transfer
reactions, where only small-size valence Fock state configurations
enter the hard scattering amplitude, both the initial and final
state interactions of the hadron states become negligible. 
Color Transparency can be measured though the nuclear dependence 
of
totally diffractive vector meson production 
$d\sigma/dt(\gamma^*A\to
V A).$ For large photon virtualities (or for heavy vector
quarkonium), the small color dipole moment of the vector system
implies minimal absorption.  Thus, remarkably, QCD predicts that the
forward amplitude $\gamma^* A \to V A$ at $t \to 0$ is nearly 
linear
in $A$.  One is also sensitive to corrections from the nonlinear
$A$-dependence of the nearly forward matrix element that couples 
two
gluons to the nucleus, which is closely related to the nuclear
dependence of the gluon structure function of the nucleus.\cite{BGMFS} 
The integral of the diffractive cross section over the forward peak
is thus predicted to scale approximately as $A^2/R_A^2 \sim
A^{4/3}.$  Evidence for color transparency in quasi-elastic $\rho$
leptoproduction $\gamma^* A \to \rho^0 N (A-1)$ has recently been
reported by the E665 experiment at Fermilab \cite{e665} for both
nuclear coherent and incoherent reactions. A test could also be
carried out at very small $t_{\rm min}$ at HERA, and would provide 
a
striking test of QCD in exclusive nuclear reactions. There is also
evidence for QCD ``color transparency" in quasi-elastic $p p$
scattering in nuclei.\cite{heppelmann90}  In contrast to color
transparency, Fock states with large-scale color configurations
interact strongly and with high particle number production.
\cite{bbfhs93}

{\it Hidden Color }
The deuteron form factor at high $Q^2$ is sensitive to wavefunction
configurations where all six quarks overlap within an impact
separation $b_{\perp i} < {\cal O} (1/Q);$ the leading power-law
fall off predicted by QCD is $F_d(Q^2) = f(\alpha_s(Q^2))/(Q^2)^5$,
where, asymptotically, $f(\alpha_s(Q^2))\propto
\alpha_s(Q^2)^{5+2\gamma}$.\cite{bc76}  The derivation of the
evolution equation for the deuteron distribution amplitude and its
leading anomalous dimension $\gamma$ is given in Ref. \cite{bjl83}  
In general, the six-quark wavefunction of a deuteron
is a mixture of five different color-singlet states. The dominant
color configuration at large distances corresponds to the usual
proton-neutron bound state. However at small impact space
separation, all five Fock color-singlet components eventually
acquire equal weight, \ie, the deuteron wavefunction evolves to
80\%\ ``hidden color.''  The relatively large normalization of the
deuteron form factor observed at large $Q^2$ points to sizable
hidden color contributions.\cite{fhzxx}

{\it Spin-Spin Correlations in Nucleon-Nucleon
Scattering and the Charm \hfill\break Threshold }
One of the most striking anomalies in elastic proton-proton
scattering is the large spin correlation $A_{NN}$ observed at large
angles.\cite{krisch92}   At $\sqrt s \simeq 5 $ GeV, the rate for
scattering with incident proton spins parallel and normal to the
scattering plane is four times larger than that for scattering with
anti-parallel polarization. This strong polarization correlation can
be attributed to the onset of charm production in the intermediate
state at this energy.\cite{bdt88}  The intermediate state $\vert u
u d u u d c \bar c \rangle$ has odd intrinsic parity and couples to
the $J=S=1$ initial state, thus strongly enhancing scattering when
the incident projectile and target protons have their spins parallel
and normal to the scattering plane.  The charm threshold can also
explain the anomalous change in color transparency observed at the
same energy in quasi-elastic $ p p$ scattering. A crucial test is
the observation of open charm production near threshold with a 
cross
section of order of $1 \mu$b.

{\it The QCD Van Der Waals Potential and
Nuclear Bound Quarkonium }
The simplest manifestation of the nuclear force is the interaction
between two heavy quarkonium states, such as the $\Upsilon (b \bar
b)$ and the $J/\psi(c \bar c)$. Since there are no valence quarks in
common, the dominant color-singlet interaction arises simply from
the exchange of two or more gluons. In principle, one could measure
the interactions of such systems by producing pairs of quarkonia in
high energy hadron collisions. The same fundamental QCD van der
Waals potential also dominates the interactions of heavy quarkonia
with ordinary hadrons and nuclei. The small size of the $Q \overline
Q$ bound state relative to the much larger hadron allows a
systematic expansion of the gluonic potential using the operator
product expansion.\cite{manoharxx}  The coupling of the scalar part
of the interaction to large-size hadrons is rigorously normalized to
the mass of the state via the trace anomaly. This scalar attractive
potential dominates the interactions at low relative velocity. In
this way one establishes that the nuclear force between heavy
quarkonia and ordinary nuclei is attractive and sufficiently strong
to produce nuclear-bound quarkonium.\cite{manoharxx,btsxx}  Recently, Miller and
I have shown that the corrections to the gluon exchange potential from meson
exchange contributions are relatively negligible, and we show how deuteron
targets can be used to measure the $J/\psi$-nucleon cross section.\cite{Miller}   
Navarra and
I have shown that exclusive decays of $B$ mesons at $B$ factories such as the
$B  \to J/\psi \bar p \Lambda $ can provide a sensitive search tool for finding
possible $J/\psi$-baryon resonances.\cite{Navarra}

\section {Commensurate Scale Relations}

A critical problem in making reliable predictions in quantum
chromodynamics is how to deal with the dependence of the truncated
perturbative series  on the choice of renormalization scale and
scheme.  For processes
where only the leading and next-to-leading order predictions are
known,  the theoretical uncertainties from the choice of
renormalization scale and scheme are often much larger than the
experimental uncertainties. 
The uncertainties introduced by the conventions in the
renormalization procedure are amplified in processes involving more
than one physical scale such as jet observables and semi-inclusive
reactions. In the case of jet production at $e^+e^-$ colliders, the
jet fractions depend both on the total center of mass energy $s$
and the jet resolution parameter $y$ (which gives an upperbound $y
s$ to the invariant mass squared of each individual jet). different scale-setting
strategies can lead to very different behaviors for the
renormalization scale in the small $y$ region. In the case of QCD
predictions for exclusive processes such as the decay of heavy
hadrons to specific channels and baryon form factors at large
momentum transfer, the scale ambiguities for the underlying
quark-gluon subprocesses are even more acute since the coupling
constant $\alpha_s(\mu)$ enters at a high power.  Furthermore, since
the external momenta entering an exclusive reaction are partitioned
among the many propagators of the underlying hard-scattering
amplitude, the physical scales that control these processes are
inevitably much softer than the overall momentum transfer.

The renormalization scale ambiguity problem can be resolved if one can
optimize the choices of scale and scheme according to some sensible
criteria. In the BLM procedure \cite{BLM}, the renormalization scales are chosen
such that all vacuum polarization effects from the QCD $\beta$
function are re-summed into the running couplings.  The coefficients
of the perturbative series are thus identical to the perturbative
coefficients of the corresponding conformally invariant theory with
$\beta=0.$ The BLM method has the important advantage of
``pre-summing" the large and strongly divergent terms in the PQCD
series which grow as $n!  (\alpha_s \beta_0 )^n$, {\em i.e.}, the
infrared renormalons associated with coupling constant renormalization.
\cite{Mueller,BallBenekeBraun}  Furthermore, the renormalization
scales $Q^*$ in the BLM method are physical in the sense that they
reflect the mean virtuality of the gluon propagators.
\cite{BLM,BallBenekeBraun,LepageMackenzie,Neubert}  In fact, in the
$\alpha_V(Q)$ scheme, where the QCD coupling is defined from the heavy
quark potential, the renormalization scale is by definition the
momentum transfer caused by the gluon.

A basic principle of renormalization theory is the requirement that
relations between physical observables must be independent of
renormalization scale and scheme conventions to any fixed order of
perturbation theory.\cite{StueckelbergPeterman}  In this section, I
shall discuss high precision perturbative predictions which have no
scale or scheme ambiguities.  These predictions, called
``Commensurate Scale Relations," \cite{BrodskyLu} are valid for any
renormalizable quantum field theory, and thus may provide a uniform
perturbative analysis of the electroweak and strong sectors of the
Standard Model.

Commensurate scale relations relate observables to observables, and
thus are independent of theoretical conventions such as choice of
intermediate renormalization scheme. The scales of the effective
charges that appear in commensurate scale relations are fixed by the
requirement that the couplings sum all of the effects of the
nonzero $\beta$ function, as in the BLM method.\cite{BLM} The
coefficients in the perturbative expansions in the commensurate
scale relations are thus identical to those of a corresponding
conformally-invariant theory with $\beta=0.$

A helpful tool and notation for relating physical quantities is the
effective charge. Any perturbatively calculable physical quantity
can be used to define an effective
charge \cite{Grunberg,DharGupta,GuptaShirkovTarasov} by incorporating
the entire radiative correction into its definition. An important
result is that all effective charges $\alpha_A(Q)$ satisfy the
Gell-Mann-Low renormalization group equation with the same $\beta_0$
and $\beta_1;$ different schemes or effective charges only differ
through the third and higher coefficients of the $\beta$ function.
Thus, any effective charge can be used as a reference running
coupling constant in QCD to define the renormalization procedure.
More generally, each effective charge or renormalization scheme,
including $\overline{\rm MS}$, is a special case of the universal
coupling function $\alpha(Q, \beta_n)$.
\cite{StueckelbergPeterman,BLM}  Peterman and St\"uckelberg
have shown \cite{StueckelbergPeterman} that all effective charges
are related to each other through a set of evolution equations in
the scheme parameters $\beta_n.$

For example, consider the entire radiative corrections to the
annihilation cross section expressed as the ``effective charge"
$\alpha_R(Q)$ where $Q=\sqrt s$:
\begin{equation}
R(Q) \equiv 3 \sum_f Q_f^2 \left[1+{\alpha_R(Q) \over \pi}
\right].
\end{equation}
Similarly, we can define the entire radiative correction to the
Bjorken sum rule as the effective charge $\alpha_{g_1}(Q)$
where $Q$
is the lepton momentum transfer:
\begin{equation}
\int_0^1
d x \left[
   g_1^{ep}(x,Q^2) - g_1^{en}(x,Q^2) \right]
   \equiv {1\over 3} \left|g_A \over g_V \right|
   \left[ 1- {\alpha_{g_1}(Q) \over \pi} \right] .
\end{equation}

The commensurate scale relations connecting the effective charges
for observables $A$ and $B$ have the form 
\begin{equation}
\alpha_A(Q_A) =
\alpha_B(Q_B) \left(1 + r_{A/B} {\alpha_B\over \pi} +\cdots\right)\ ,
\end{equation}
where the coefficient $r_{A/B}$ is independent of the number of
flavors $n_F$ contributing to coupling constant renormalization. 
We calculate the coefficients in the next section.
The ratio of scales $\lambda_{A/B} = Q_A/Q_B$ is unique at leading
order and guarantees that the observables $A$ and $B$ pass through
new quark thresholds at the same physical scale.  One also can show
that the commensurate scales satisfy the transitivity rule
$\lambda_{A/B} = \lambda_{A/C} \lambda_{C/B},$ which is the
renormalization group property which ensures that predictions in
PQCD are independent of the choice of an intermediate
renormalization scheme $C.$ In particular, scale-fixed predictions
can be made without reference to theoretically-constructed
renormalization schemes such as $MSb.$ QCD can thus be tested in a
new and precise way by checking that the observables track both in
their relative normalization and in their commensurate scale
dependence.

A scale-fixed relation between any two physical observables $A$ and
$B$ can be derived by applying BLM scale-fixing to their respective
perturbative predictions in, say, the $\overline {MS}$ scheme, and
then algebraically eliminating $\alpha_{\overline {MS}}.$ The choice
of the BLM scale ensures that the resulting commensurate scale
relation between $A$ and $B$ is independent of the choice of the
intermediate renormalization scheme.\cite{BLM}  Thus, using this
formalism, one can relate any perturbatively calculable observable,
such as the annihilation ratio $R_{e^+ e^-}$, the heavy quark
potential, and the radiative corrections to structure function sum
rules to each other without any renormalization scale or scheme
ambiguity.\cite{BrodskyLu}  Commensurate scale relations can also be
applied in grand unified theories to make scale and scheme invariant
predictions which relate physical observables in different sectors
of the theory.

Scales that appear in commensurate scale relations are physical
since they reflect the mean virtuality of the gluons in the
underlying hard subprocess.\cite{BLM,Neubert}  As emphasized by
Mueller,\cite{Mueller} commensurate scale relations isolate the
effect of infrared renormalons associated with the nonzero $\beta$
function. The usual factorial growth of the coefficients in
perturbation theory due to quark and gluon vacuum polarization
insertions is eliminated since such effects are resummed into the
running couplings. The perturbative series is thus much more
convergent.

It is interesting to compare Pad\'e resummation predictions for
single-scale perturbative QCD series in which the initial
renormalization scale choice is taken as the characteristic scale
$\mu=Q$ as well as the BLM scale $\mu=Q^*$.  One finds
\cite{BEGKS}
that the Pad\'e predictions for the summed series are in each case
independent of the initial scale choice, an indication that the
Pad\'e results are thus characteristic of the actual QCD prediction.
However, the BLM scale generally produces a faster convergence to
the complete sum than the conventional scale choice.  This can be
understood by the fact that the BLM scale choice immediately sums
into the coupling all repeated vacuum polarization insertions to all
orders, thus eliminating the large $(\beta_0\alpha_s)^n$ terms in
the series as well as the $n!$ growth characteristic of the infrared
renormalon structure of PQCD.\cite{tHooft,Mueller}

\section {The Generalized Crewther Relation}

In 1972 Crewther \cite{Crewther} derived a  remarkable consequence
of the operator product expansion for conformally-invariant gauge
theory.  Crewther's relation has the form
\begin{equation}
3 S = K R'
\end{equation}
where $S$ is the value of the anomaly controlling $\pi^0 \to \gamma
\gamma$ decay, $K$ is the value of the Bjorken sum rule in polarized
deep inelastic scattering, and $R'$ is the isovector part of the
annihilation cross section ratio $\sigma(e^+ e^- \to
$hadrons)/$\sigma(e^+ e^- \to \mu^+ \mu^-)$. Since $S$ is unaffected
by QCD radiative corrections,\cite{Bardeen}  Crewther's relation
requires that the QCD radiative corrections to $R_{e^+ e^-}$ exactly
cancel the radiative corrections to the Bjorken sum rule order by
order in perturbation theory.

However,  Crewther's relation is only valid in the case of
conformally-invariant gauge theory, {\it i.e.} when the coupling
$\alpha_s$ is scale invariant. However, in reality the radiative
corrections to the Bjorken sum rule and the annihilation ratio are
in general functions of different physical scales. Thus Crewther's
relation cannot be tested directly in QCD unless the effects of the
nonzero $\beta$ function for the QCD running coupling are accounted
for, and the energy  scale $\sqrt s$ in the annihilation cross
section is related to the momentum transfer $Q$ in the deep
inelastic sum rules. Recently Broadhurst and Kataev
\cite{BroadhurstKataev} have explicitly calculated the radiative
corrections to the Crewther relation and have demonstrated
explicitly that the corrections are proportional to the QCD $\beta$
function.

We can use the known expressions to three loops
\cite{LarinVermaseren,GorishnyKataevLarin,SurguladzeSamuel} in
$\bar{\rm MS}$ scheme and choose the leading-order and
next-to-leading scales $Q^*$ and $Q^{**}$ to re-sum all quark and
gluon vacuum polarization corrections into the running couplings.
The values of these scales are the physical values of the energies
or momentum transfers which ensure that the radiative corrections to
each observable passes through the heavy quark thresholds at their
respective commensurate physical scales.  The final result
connecting the effective charges (see Section 1) is remarkably
simple:
\begin{equation}
{\alpha_{g_1}(Q) \over \pi} = {\alpha_R(Q^*) \over \pi} -
\left( {\alpha_R(Q^{**}) \over \pi} \right)^2
+ \left( {\alpha_R(Q^{f})\over \pi} \right)^3 + \cdots .
\label{AlphaG1AlphaRAfterBLMThreeFlavors}
\end{equation}
The coefficients in the series (aside for a factor of $C_F,$ which
can be absorbed in the definition of $\alpha_s$) are actually
independent of color and are the same in Abelian, non Abelian, and
conformal gauge theory.  The non-Abelian structure of the theory is
reflected in the scales $Q^*$ and $Q^{**}.$ Note that the a
calculational device; it simply serves as an intermediary between
observables and does not appear in the final relation
(\ref{AlphaG1AlphaRAfterBLMThreeFlavors}). This is equivalent to the
group property defined by Peterman and St\"uckelberg
\cite{StueckelbergPeterman} which ensures that predictions in PQCD
are independent of the choice of an intermediate renormalization
scheme. (The renormalization group method was developed by Gell-Mann
and Low \cite{GellMannLow} and by Bogoliubov and
Shirkov.\cite{BogoliubovShirkov})

The connection between the effective charges of
observables given by Eq. (\ref{AlphaG1AlphaRAfterBLMThreeFlavors}) 
is a prime example of a ``commensurate scale relation" (CSR).  A fundamental test of QCD
will be to verify empirically that the related observables track in both
normalization and shape as given by the CSR.  The commensurate scale
relations thus provide fundamental tests of QCD which can be made
increasingly precise and independent of the choice of
renormalization scheme or other theoretical convention. More
generally, the CSR between sets of physical observables
automatically satisfy the transitivity and symmetry properties
\cite{BrodskyLuSelfconsistency} of the scale transformations of the
renormalization ``group" as originally defined by Peterman and
St\"uckelberg.\cite{StueckelbergPeterman}  The predicted relation
between observables must be independent of the order one makes
substitutions; {\it i.e.} the algebraic path one takes to relate the
observables.

The relation between scales in the CSR is consistent with the BLM
scale-fixing procedure \cite{BLM} in which the scale is chosen such
that all terms arising from the QCD $\beta-$function are resummed
into the coupling.  Note that this also implies that the
coefficients in the perturbation CSR expansions are independent of
the number of quark flavors $f$ renormalizing the gluon propagators.
This prescription ensures that, as in quantum electrodynamics,
vacuum polarization contributions due to fermion pairs are all
incorporated into the coupling $\alpha(\mu)$ rather than the
coefficients. The coefficients in the perturbative expansion using
BLM scale-fixing are the same as those of the corresponding
conformally invariant theory with $\beta=0.$ In practice, the
conformal limit is defined by $\beta_0, \beta_1 \to 0$, and can be
reached, for instance, by adding enough spin-half and scalar quarks
as in $N=4$ supersymmetric QCD.  Since all the running coupling
effects have been absorbed into the renormalization scales, the  BLM
scale-setting method correctly reproduces the perturbation theory
coefficients of the conformally invariant theory in the $\beta \to
0$ limit.

The commensurate scale relation between $\alpha_{g_1}$ and
$\alpha_R$ given by Eq. (\ref{AlphaG1AlphaRAfterBLMThreeFlavors})
implies that the radiative corrections to the annihilation cross
section and the Bjorken (or Gross-Llewellyn Smith) sum rule cancel
at their commensurate scales.  The relations between the physical
cross sections can be written in the forms:
\begin{equation}
{R_{e^+ e^-}(s)\over 3\sum e^2_q} ~
{\int^1_0 dx  g_1^p(x,Q^2)-g_1^n(x,Q^2) \over {1\over 3}
g_A/g_V}
= 1 - \Delta \beta_0 \widehat a^3
\end{equation}
and
\begin{equation}
{R_{e^+ e^-}(s)\over 3\sum e^2_q} ~
{\int^1_0 dx  F_3^{\nu p}(x,Q^2) + F_3^{\bar\nu p}(x,Q^2)
\over 6} = 1 - \Delta \beta_0 \widehat a^3,
\label{eq10a}
\end{equation}
provided that the annihilation energy in $R_{e^+ e^-}(s)$ and the
momentum transfer $Q$ appearing in the deep inelastic structure
functions are commensurate at NLO: $\sqrt s = Q^* = Q \exp [{7\over
4}- 2\zeta_3 + ({11\over 96} +{7\over 3} \zeta_3 - 2\zeta^2_3 -
{\pi^2\over 24})\beta_0 \widehat a(Q)]$. The light-by-light
correction to the CSR for the Bjorken sum rule vanishes for three
flavors.  The term $\Delta \beta_0 \widehat a^3$ with $\Delta = \ell
n\, (Q^{**}/Q^*)$ is the third-order correction arising from the
difference between $Q^{**}$ and $Q^*$; in practice this correction
is negligible: for a typical value $\widehat a = \alpha_R(Q)/ \pi =
0.14,$ $\Delta \beta_0 \widehat a^3 = 0.007.$ Thus at the magic
energy $\sqrt s = Q^*$, the radiative corrections to the Bjorken and
GLLS sum rules almost precisely cancel the radiative corrections to
the annihilation cross section.  This allows a practical test and
extension of the Crewther relation to nonconformal QCD.

As an initial test of Eq. (\ref{eq10a}), we can compare the CCFR
measurement \cite{CCFR} of the Gross-Llewellyn Smith sum rule
$1-\widehat\alpha_{F_3} = {1 \over 6}\int^1_0 dx [F_3^{\nu p}(x,Q^2)
+ F_3^{\bar\nu p}(x,Q^2)] = {1\over 3} ( 2.5 \pm 0.13 )$ at
$Q^2 = 3$ GeV$^2$ and the parameterization of the annihilation data
\cite{MattinglyStevenson} $1 + \widehat\alpha_R = R_{e^+
e^-}(s)/3\sum e^2_q = 1.20.$ at the commensurate scale $\sqrt s =
Q^*= 0.38\, Q = 0.66$ GeV. The product is $(1 + \widehat\alpha_R)(1
-\widehat\alpha_{F_3})=1.00 \pm 0.04$, which is a highly nontrivial
check of the theory at very low physical scales. More recently, the
E143 \cite{E143} experiment at SLAC has reported a new value for the
Bjorken sum rule at $Q^2= 3\ $GeV$^2$: $\Gamma_1^p - \Gamma_1^n =
0.163 \pm 0.010 ({\rm stat}) \pm 0.016 ({\rm syst}).$ The Crewther
product in this case is also consistent with QCD: $(1 +
\widehat\alpha_R)(1 -\widehat\alpha_{g_1})=0.93 \pm 0.11.$

In a paper with Gabadadze, Kataev and Lu \cite {BGKL} we show that
it is also possible and convenient to choose one unique mean scale
$\bar Q^*$ in $\alpha_R(Q)$ so that the perturbative expansion
will also reproduce the coefficients of the geometric progression.
The possibility of using a single scale in the generalization of the
BLM prescription beyond the next-to-leading order (NLO) was first
considered by Grunberg and Kataev.\cite{GrunKat}  The new single-scale
Crewther relation has the form:
\begin{equation}
\widehat{\alpha}_{g_1}(Q)=\widehat{\alpha}_R(\bar
Q^*)-
\widehat{\alpha}_R^2(\bar
Q^*)+\widehat{\alpha}_R^3(\bar Q^*)
+ \cdots,
\end{equation}

The generalized Crewther relation  provides an important test of
QCD. Since the Crewther formula written in the form of the CSR
relates one observable to another observable, the predictions are
independent of theoretical conventions, such as the choice of
renormalization scheme. It is clearly  very interesting to test
these fundamental self-consistency relations between the polarized
Bjorken sum rule or the Gross-Llewellyn Smith sum rule and the
$e^+e^-$-annihilation $R$-ratio. Present data are consistent with
the generalized Crewther relations, but measurements at higher
precision and energies will be needed to decisively test these
fundamental connections in QCD.

It is worthwhile to point out that commensurate scale relations
are derived within the framework of perturbation theory in leading
twist and do not involve the nonperturbative contributions to the
Adler's function $D(Q^2)$\cite{SVZ} and the $R$-ratio, as well as
to  the polarized Bjorken and the Gross-Llewellyn Smith sum rules. 
\cite{Jaffe,BBK}  These nonperturbative contributions are expected
to be significant at small energies and momentum  transfer. In order
to make these contributions comparatively negligible,  one should
choose relatively large values for $s$ and $Q^2$.   In order to put
the analysis of the experimental data for lower energies on more
solid ground, it will be necessary to understand whether there exist
any Crewther-type relations between nonperturbative order
$O(1/Q^4)$-corrections to the Adler's $D$-function \cite{SVZ} and
the order $O(1/Q^2)$ higher twist contributions to the
deep-inelastic sum rules.\cite{Jaffe,BBK}

Commensurate scale relations such as the generalized Crewther
relation discussed here open up additional possibilities for testing
QCD.  One can compare two observables by checking that their
effective charges agree both in normalization and in their scale
dependence.  The ratio of leading-order commensurate scales
$\lambda_{A/B}$ is fixed uniquely: it ensures that both observables
$A$ and $B$ pass through heavy quark thresholds at precisely the
same physical point. The same procedure can be applied to
multi-scale problems; in general, the commensurate scales $Q^*,
Q^{**}$, etc. will depend on all of the available scales.

The coefficients in a  CSR are identical to the coefficients in a
conformal theory where explicit renormalon behavior does not appear.
It is thus reasonable to expect that the series expansions appearing
in the CSR are convergent when one relates finite observables to
each other. Thus commensurate scale relations between observables
allow tests of perturbative QCD with higher and higher precision as
the perturbative expansion grows.

\section{Renormalization Scale Fixing In Exclusive Processes}

As we have noted, perturbative QCD can be used to analyze a number of exclusive
processes involving large momentum transfers, including the decay of
heavy hadrons to specific channels such as $B \to \pi \pi$ and
$\Upsilon \to p \bar p$, baryon form factors at large $t$, and fixed
$\theta_{c.m.}$ hadronic scattering amplitudes such as $\gamma p \to
\pi^+ n$ at high energies.\cite{BLReview} As in the case of inclusive reactions,
factorization theorems for exclusive processes
\cite{BrodskyLepage,EfremovRad} allow the analytic separation
of the perturbatively-calculable short-distance contributions from the
long-distance nonperturbative dynamics associated with hadronic
binding.

The scale ambiguities for the underlying quark-gluon subprocesses are
particularly acute in the case of QCD predictions for exclusive
processes, since the running coupling $\alpha_s$ enters at a high
power. Furthermore, since each external momentum entering an exclusive
reaction is partitioned among the many propagators of the underlying
hard-scattering amplitude, the physical scales that control these
processes are inevitably much softer than the overall momentum
transfer. Exclusive process phenomenology is further complicated by
the fact that the scales of the running couplings in the
hard-scattering amplitude depend themselves on the shape of the
hadronic wavefunctions.

In this section  we will discuss the application of the BLM method to fix the
renormalization scale of the QCD coupling in exclusive hadronic amplitudes such
as the pion form factor, the photon-to-pion transition form factor and
$\gamma \gamma \rightarrow \pi^+ \pi^-$ at large momentum transfer.
Renormalization-scheme-independent commensurate scale relations will
be established which connect the hard scattering subprocess amplitudes
that control these exclusive processes to other QCD observables such
as the heavy quark potential and the electron-positron annihilation
cross section.  Because the renormalization scale is small, we will
argue that the effective coupling is nearly constant, thus accounting
for the nominal (dimensional counting) scaling behavior 
\cite{BrodskyFarrar} of the data.\cite{JiSillLombard,JiAmiri}

The heavy-quark potential $V(Q^2)$ can be identified as the
two-particle-irreducible scattering amplitude of test charges, {\em
i.e.}, the scattering of an infinitely heavy quark and antiquark at
momentum transfer $t = -Q^2.$ The relation
\begin{equation}
V(Q^2) = -  {4 \pi C_F \alpha_V(Q^2)\over Q^2},
\end{equation}
with $C_F=(N_C^2-1)/2 N_C=4/3$, then defines the effective charge
$\alpha_V(Q).$ This coupling provides a physically-based alternative
to the usual ${\overline {MS}}$ scheme.  Recent lattice gauge
calculations have provided strong constraints on the normalization and
shape of $\alpha_V(Q^2)$.

As in the corresponding case of Abelian QED, the scale $Q$ of the
coupling $\alpha_V(Q)$ is identified with the exchanged momentum.  All
vacuum polarization corrections due to fermion pairs are incorporated
in terms of the usual vacuum polarization kernels defined in terms of
physical mass thresholds.  The first two terms $\beta_0 = 11 - 2
n_f/3$ and $\beta_1 = 102 - 38n_f/3$ in the expansion of the $\beta$
function defined from the logarithmic derivative of $\alpha_V(Q)$ are
universal, {\em i.e.}, identical for all effective charges at $Q^2 \gg
4m_f^2$.  The coefficient $\beta_2$ for $\alpha_V$ has recently been
calculated in the $\overline {MS}$ scheme.\cite{MarkusPeter}

The scale-fixed relation between $\alpha_V$ and the conventional
$\overline {MS}$ coupling is
\begin{equation}
\alpha_{\overline {MS}}(Q)= \alpha_V(e^{5/6} Q) \left(1
+ \frac{2C_A}{3} {\alpha_V\over\pi} + \cdots\right),
\label{alpmsbar}
\end {equation}
above or below any quark mass threshold.  The factor $e^{5/6} \simeq
0.4346$ is the ratio of commensurate scales between the two schemes to
this order.  It arises because of the convention used in defining the
modified minimal subtraction scheme. The scale in the $\overline {MS}$
scheme is thus a factor $\sim 0.4$ smaller than the physical scale.
The coefficient $2C_A/3$ in the NLO term is a feature of the
non-Abelian couplings of QCD; the same coefficient occurs even if the
theory had been conformally invariant with $\beta_0=0.$ 
The commensurate scale relation
between $\alpha_V$, as defined from the heavy quark potential, and
$\alpha_{\bar {MS}}$ provides an analytic extension of the
$\bar {MS}$ scheme in which flavor thresholds are
automatically. taken into account at their proper respective
scales.\cite{Mirabelli,BrodskyLu} The coupling $\alpha_V$ provides a natural scheme for
computing exclusive amplitudes. Once we relate form factors to
effective charges based on observables, there are no ambiguities due
to scale or scheme conventions.

The use of $\alpha_V$ as the expansion parameter with BLM scale-fixing
has also been found to be valuable in lattice gauge theory, greatly
increasing the convergence of perturbative expansions relative to
those using the bare lattice coupling.\cite{LepageMackenzie}  In fact,
new lattice calculations of the $\Upsilon$ spectrum\cite{Davies} have
been used to determine the normalization of the static heavy quark
potential and its effective charge:
\begin{equation}
\alpha_V^{(3)}(8.2~{\rm GeV}) = 0.196(3),
\label{alpv8.2}
\end{equation}
where the effective number of light flavors is $n_f = 3$. The
corresponding modified minimal subtraction coupling evolved to the $Z$
mass using Eq. (\ref{alpmsbar}) is given by
\begin{equation}
\alpha_{\overline{MS}}^{(5)}(M_Z) = 0.115(2).
\label{alpmsbarmz}
\end{equation}
This value is consistent with the world average of 0.117(5), but is
significantly more precise. These results are valid up to NLO.

\section{Hard Exclusive Two-Photon Reactions}

Exclusive two-photon processes such as $\gamma \gamma \to $ hadron
pairs and the transition form factor $\gamma^* \gamma \to $ neutral
mesons   play a unique role  in testing quantum chromodynamics because
of the simplicity of the initial state.\cite{BrodskyLepage}  At large
momentum transfer the direct point-like coupling of the photon
dominates at leading twist, leading to highly specific predictions
which depend on the shape and normalization of 
the hadron distribution
amplitudes $\phi_H(x_i,Q),$ 
the basic valence bound state wavefunctions.
The most recent exclusive two-photon process data from CLEO
\cite{Dominick} provides stringent tests of these fundamental QCD
predictions.

Exclusive processes are particularly sensitive to the unknown nonperturbative
bound state dynamics of the hadrons. However, in some important cases, the leading
power-law behavior of an exclusive amplitude at large momentum transfer can be
computed rigorously via a factorization theorem which separates the soft and hard
dynamics. The key ingredient is the factorization of the hadronic amplitude at
leading twist. As in the case of inclusive reactions, factorization
theorems for exclusive processes \cite{BrodskyLepage,EfremovRad,BLReview} allow
the analytic separation of the perturbatively-calculable short-distance
contributions from the long-distance nonperturbative dynamics
associated with hadronic binding.  
For example, the
amplitude $\gamma\gamma \rightarrow \pi^+\pi^-$ factorizes in the form
\begin{equation} 
\M_{\gamma\gamma \rightarrow \pi^+\pi^-} = \int^1_0 dx \int^1_0 dy\,
\phi_\pi(x,\widetilde Q)\, T_H(x,y,\widetilde Q)\,
\phi_\pi(y,\widetilde Q) 
\end{equation} 
where $\phi_\pi(x,\widetilde Q)$ is in the pion distribution amplitude
and contains all of the soft, nonperturbative dynamics of the pion
$q\bar q$ wavefunction integrated in relative transverse momentum up to
the separation scale $k_\perp^2 < \widetilde Q^2$, and $T_H$ is the
quark/gluon hard scattering amplitude for $\gamma\gamma \rightarrow
(q\bar q)(q\bar q)$ where the outgoing quarks are taken collinear with
their respective pion parent.  To lowest order in $\alpha_s$, the hard
scattering amplitude is linear in $\alpha_s$.  The most convenient
definition of the coupling is the effective charge $\alpha_V(Q^2)$,
defined from the potential for the scattering of two infinitely heavy
test charges, in analogy to the definition of the QED running coupling.
Another possible choice is the effective charge $\alpha_R(s)$, defined
from the QCD correction to the annihilation cross section: $R_{e^+e^-
\to {\rm hadrons}}(s) \equiv R_0 (1 + \alpha_R(s)/\pi).$ One can relate
$\alpha_V$ and $\alpha_R$ to $\alpha_{\bar{MS}}$ to NNLO using
commensurate scale relations.\cite{BrodskyLu}

The contributions from non-valence Fock states and the correction from
neglecting the transverse momentum in the subprocess amplitude from the
nonperturbative region are higher twist, {\em i.e.}, power-law
suppressed. The transverse momenta in the perturbative domain lead to
the evolution of the distribution amplitude and to
next-to-leading-order (NLO) corrections in $\alpha_s$.  The
contribution from the endpoint regions of integration, $x \sim 1$ and
$y \sim 1,$ are power-law and Sudakov suppressed and thus can only
contribute corrections at higher order in $1/Q$.\cite{BrodskyLepage}

The QCD coupling is typically evaluated at quite low scales in
exclusive processes since the momentum transfers has to be divided
among several constituents.  In the BLM procedure, the scale of the
coupling is evaluated by absorbing all vacuum polarization corrections
with the scale of the coupling or by taking the experimental value
integrating over the gluon virtuality.  Thus, in the case of the
(timelike) pion form factor the relevant scale is of order $Q^{*2} \sim
e^{-3}\M^2_{\pi\pi^-} \cong \frac{1}{20}\, \M^2_{\pi^+\pi^-}$ assuming
the asymptotic form of the pion distribution amplitude $\phi^{\rm
asympt}_\pi = \sqrt 3\, f_\pi\, x(1-x)$.  At such low scales, it is
likely that the coupling is frozen or relatively slow varying.

In the BLM procedure, the renormalization scales are chosen such that
all vacuum polarization effects from the QCD $\beta$ function are
re-summed into the running couplings.  The coefficients of the
perturbative series are thus identical to the perturbative coefficients
of the corresponding conformally invariant theory with $\beta=0.$ The
BLM method has the important advantage of ``pre-summing" the large and
strongly divergent terms in the PQCD series which grow as $n! 
(\alpha_s \beta_0 )^n$, {\em i.e.}, the infrared renormalons associated
with coupling constant renormalization.\cite{Mueller,BallBenekeBraun}
Furthermore, the renormalization scales $Q^*$ in the BLM method are
physical in the sense that they reflect the mean virtuality of the
gluon propagators.\cite{BallBenekeBraun,BLM,LepageMackenzie,Neubert} 
In fact, in the $\alpha_V(Q)$ scheme, where the QCD coupling is defined
from the heavy quark potential, the renormalization scale is by
definition the momentum transfer caused by the gluon. Because the
renormalization scale is small in the exclusive $\gamma \gamma$
processes discussed here, we will argue that the effective coupling is
nearly constant, thus accounting for the nominal scaling behavior of
the data.\cite{JiSillLombard,JiAmiri}

Ji, Pang, Robertson, and I \cite{BJPR} have recently analyzed the pion
transition form factor $F^{\gamma^*\gamma} \rightarrow \pi^0$ obtained
from $e\gamma \rightarrow e^\prime \pi^0$, the timelike pion form
obtained from $e^+e^- \rightarrow \pi^+\pi$, and the $\gamma\gamma
\rightarrow \pi^+\pi^-$ processes, all at NLO in $\alpha_V$. The
assumption of a nearly constant coupling in the hard scattering
amplitude at low scales provides an explanation for the
phenomenological success of dimensional counting rules for exclusive
processes; \ie, the power-law fall-off follows the nominal scaling of
the hard scattering amplitude $\M_{\rm had} \sim T_H\sim [p_T]^{4-n}$
where $n$ is in the total number of incident and final fields entering
$T_H$.\cite{BrodskyFarrar}

The transition form factor has now been measured up to $Q^2 < 8
$ GeV$^2$ in the tagged two-photon collisions $e \gamma \to e' \pi^0$
by the CLEO and CELLO collaborations.  In this case the amplitude has
the factorized form
\begin{equation}
F_{\gamma M}(Q^2)= {4 \over \sqrt 3}\int^1_0 dx \phi_M(x,Q^2)
T^H_{\gamma \to M}(x,Q^2) ,
\label{transitionformfactor}
\end{equation}
where the hard scattering amplitude for $\gamma \gamma^* \to q \bar q$
is
\begin{equation}
T^H_{\gamma M}(x,Q^2) = {1\over (1-x) Q^2}\left(1 +
{\cal O}(\alpha_s)\right).
\label{transitionhardscattering}
\end{equation}
The leading QCD corrections have been computed by Braaten
\cite{Braaten} and Dittes and Radyushkin \cite{DittesRadyushkin}; 
however, the NLO corrections are necessary to fix the
BLM scale at LO.  Thus it is not yet possible to rigorously determine
the BLM scale for this quantity.  We shall here assume that this scale
is the same as that occurring in the prediction for $F_\pi$.  For the
asymptotic distribution amplitude we thus predict
\begin{equation}
Q^2 F_{\gamma \pi}(Q^2)= 2 f_\pi \left(1 - {5\over3}
{\alpha_V(Q^*)\over \pi}\right).
\label{qsquaretransition}
\end{equation}
As we shall see, given the phenomenological form of $\alpha_V$ we
employ (discussed below), this result is not terribly sensitive to the
precise value of the scale.

An important prediction resulting from the factorized form of these
results is that the normalization of the ratio
\begin{eqnarray}
R_\pi(Q^2) &\equiv & \frac{F_\pi (Q^2)}{4 \pi Q^2 |F_{\pi
\gamma}(Q^2)|^2}
\label{Rpidef}\\
&=& \alpha_{\overline{MS}}(e^{-14/6}Q)\left(1-0.56
{\alpha_{\overline{MS}} \over\pi} \right)
\label{Rpistart}\\
&=& \alpha_V(e^{-3/2}Q)\left(1+1.43 {\alpha_V\over\pi} \right)
\label{RpiV}\\
&=& \alpha_R(e^{5/12-{2\zeta_3}}Q) \left(1-
0.65 {\alpha_R\over\pi} \right)
\label{Rpiend}
\end{eqnarray}
is formally independent of the form of the pion distribution amplitude.
The $\alpha_{\overline{MS}}$ correction follows from combined
references.\cite{Braaten,DittesRadyushkin,Field}  The next-to-leading
correction given here assumes the asymptotic distribution amplitude.

We emphasize that when we relate $R_\pi$ to $\alpha_V$ we relate
observable to observable and thus there is no scheme ambiguity. 
Furthermore, effective charges such as $\alpha_V$  are defined from
physical observables and thus must be finite even at low momenta. A
number of proposals have been suggested for the form of the QCD
coupling in the low-momentum regime.  For example, Petronzio and Parisi
\cite{PetronzioParisi} have argued that the coupling must freeze at low
momentum transfer in order that perturbative QCD loop integrations be
well defined.  Mattingly and Stevenson \cite{MattinglyStevenson} have
incorporated such behavior into their parameterizations of $\alpha_R$
at low scales.  Gribov \cite{Gribov} has presented novel dynamical
arguments related to the nature of confinement for a fixed coupling at
low scales.  Zerwas\cite{Zerwas} has noted the heavy quark potential
must saturate to a Yukawa form since the light-quark production
processes will screen the linear confining potential at large
distances.  Cornwall \cite{Cornwall} and others
\cite{DonnachieLandshoff,GayDucatietal} have argued that the gluon
propagator will acquire an effective gluon mass $m_g$ from
nonperturbative dynamics, which again will regulate the form of the
effective couplings at low momentum. We shall adopt the simple
parameterization
\begin{equation}
\alpha_V(Q) = {4 \pi \over {\beta_0 \ln \left({{Q^2 + 4m_g^2}\over
\Lambda^2_V}\right)}} ,
\label{frozencoupling}
\end{equation}
which effectively freezes the $\alpha_V$ effective charge to a finite
value for $Q^2 \leq 4m_g^2.$

\begin{figure}[htbp]
\begin{center}
\hbox to 3in{\input{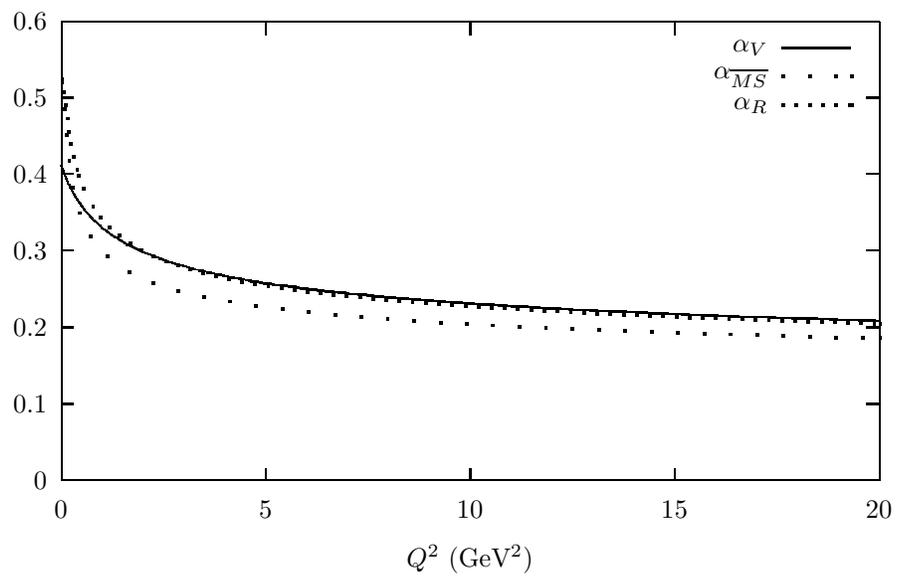}}
\end{center}
\caption[*]{The coupling function $\alpha_V(Q^2)$ as given in
Eq. (\ref{frozencoupling}).  Also shown are the corresponding
predictions for $\alpha_{\overline{MS}}$ and $\alpha_R$ following from
the NLO commensurate scale relations.}
\label{couplings}
\end{figure}

We can use the nonrelativistic heavy quark lattice results
\cite{Davies,Sloan} to fix the parameters.  A fit to the lattice data
of the above parameterization gives $\Lambda_V = 0.16$ GeV if we use
the well-known momentum-dependent $n_f$.\cite{ShirkovMikhailov}
Furthermore, the value $m^2_g=0.19$ GeV$^2$ gives consistency with the
frozen value of $\alpha_R$ advocated by Mattingly and Stevenson.
\cite{MattinglyStevenson}  Their parameterization implies the
approximate constraint $\alpha_R(Q)/\pi \simeq 0.27$ for $Q= \sqrt s <
0.3$ GeV, which leads to $\alpha_V(0.5~{\rm GeV}) \simeq 0.37$ using
the NLO commensurate scale relation between $\alpha_V$ and $\alpha_R$.
The resulting form for $\alpha_V$ is shown in Fig. \ref{couplings}. The
corresponding predictions for $\alpha_R$ and $\alpha_{\overline {MS}}$
using the CSRs at NLO are also shown.  Note that for low $Q^2$ the
couplings, although frozen, are large.  Thus the NLO and higher-order
terms in the CSRs are large, and inverting them perturbatively to NLO
does not give accurate results at low scales.  In addition,
higher-twist contributions to $\alpha_V$ and $\alpha_R$, which are not
reflected in the CSR relating them, may be expected to be important for
low $Q^2$.\cite{Braun}

\vspace{.5cm}

It is clear that exclusive processes such as the photon-to-pion
transition form factors can provide a valuable window for determining
the magnitude and the shape of the effective charges at quite low
momentum transfers.  In particular, we can check consistency with the
$\alpha_V$ prediction from lattice gauge theory.  A complimentary
method for determining $\alpha_V$ at low momentum is to use the angular
anisotropy of $e^+ e^- \to Q \overline Q$ at the heavy quark thresholds. 
\cite{BrodskyKuhnHoangTuebner} It should be emphasized that the
parameterization (\ref{frozencoupling}) is just an approximate form.
The actual behavior of $\alpha_V(Q^2)$ at low $Q^2$ is one of the key
uncertainties in QCD phenomenology.

As we have emphasized, exclusive processes are sensitive to the
magnitude and shape of the QCD couplings at quite low momentum
transfer: $Q_V^{*2} \simeq e^{-3} Q^2 \simeq Q^2/20$ and $Q_R^{*2}
\simeq Q^2/50$.\cite{LewellynIsgur}  The fact that the data for
exclusive processes such as form factors, two photon processes such as
$\gamma \gamma \to \pi^+ \pi^-,$ and photoproduction at fixed
$\theta_{c.m.}$ are consistent with the nominal scaling of the
leading-twist QCD predictions (dimensional counting) at momentum
transfers $Q$ up to the order of a few GeV can be immediately
understood if the effective charges $\alpha_V$ and $\alpha_R$ are
slowly varying at low momentum.  The scaling of the exclusive amplitude
then follows that of the subprocess amplitude $T_H$ with effectively
fixed coupling. 
Donnachie and Landshoff \cite{DL} have also argued that a frozen coupling is needed to explain the observed $t^{-8}$ scaling of $d\sigma/dt\,(pp\rightarrow pp)$ at large $s \gg -t$.  Note also that the Sudakov effect of the end point
region is the exponential of a double log series if the coupling is
frozen, and thus is strong.

\vspace{.5cm}
\begin{figure}[htbp]
\begin{center}
\hbox to 3in{% GNUPLOT: LaTeX picture
\setlength{\unitlength}{0.240900pt}
\ifx\plotpoint\undefined\newsavebox{\plotpoint}\fi
\sbox{\plotpoint}{\rule[-0.200pt]{0.400pt}{0.400pt}}%
\begin{picture}(1500,900)(0,0)
\font\gnuplot=cmr10 at 10pt
\gnuplot
\sbox{\plotpoint}{\rule[-0.200pt]{0.400pt}{0.400pt}}%
\put(219.0,134.0){\rule[-0.200pt]{4.818pt}{0.400pt}}
\put(197,134){\makebox(0,0)[r]{0}}
\put(1416.0,134.0){\rule[-0.200pt]{4.818pt}{0.400pt}}
\put(219.0,278.0){\rule[-0.200pt]{4.818pt}{0.400pt}}
\put(197,278){\makebox(0,0)[r]{0.05}}
\put(1416.0,278.0){\rule[-0.200pt]{4.818pt}{0.400pt}}
\put(219.0,422.0){\rule[-0.200pt]{4.818pt}{0.400pt}}
\put(197,422){\makebox(0,0)[r]{0.1}}
\put(1416.0,422.0){\rule[-0.200pt]{4.818pt}{0.400pt}}
\put(219.0,567.0){\rule[-0.200pt]{4.818pt}{0.400pt}}
\put(197,567){\makebox(0,0)[r]{0.15}}
\put(1416.0,567.0){\rule[-0.200pt]{4.818pt}{0.400pt}}
\put(219.0,711.0){\rule[-0.200pt]{4.818pt}{0.400pt}}
\put(197,711){\makebox(0,0)[r]{0.2}}
\put(1416.0,711.0){\rule[-0.200pt]{4.818pt}{0.400pt}}
\put(219.0,855.0){\rule[-0.200pt]{4.818pt}{0.400pt}}
\put(197,855){\makebox(0,0)[r]{0.25}}
\put(1416.0,855.0){\rule[-0.200pt]{4.818pt}{0.400pt}}
\put(219.0,134.0){\rule[-0.200pt]{0.400pt}{4.818pt}}
\put(219,89){\makebox(0,0){0}}
\put(219.0,835.0){\rule[-0.200pt]{0.400pt}{4.818pt}}
\put(462.0,134.0){\rule[-0.200pt]{0.400pt}{4.818pt}}
\put(462,89){\makebox(0,0){2}}
\put(462.0,835.0){\rule[-0.200pt]{0.400pt}{4.818pt}}
\put(706.0,134.0){\rule[-0.200pt]{0.400pt}{4.818pt}}
\put(706,89){\makebox(0,0){4}}
\put(706.0,835.0){\rule[-0.200pt]{0.400pt}{4.818pt}}
\put(949.0,134.0){\rule[-0.200pt]{0.400pt}{4.818pt}}
\put(949,89){\makebox(0,0){6}}
\put(949.0,835.0){\rule[-0.200pt]{0.400pt}{4.818pt}}
\put(1193.0,134.0){\rule[-0.200pt]{0.400pt}{4.818pt}}
\put(1193,89){\makebox(0,0){8}}
\put(1193.0,835.0){\rule[-0.200pt]{0.400pt}{4.818pt}}
\put(1436.0,134.0){\rule[-0.200pt]{0.400pt}{4.818pt}}
\put(1436,89){\makebox(0,0){10}}
\put(1436.0,835.0){\rule[-0.200pt]{0.400pt}{4.818pt}}
\put(219.0,134.0){\rule[-0.200pt]{293.175pt}{0.400pt}}
\put(1436.0,134.0){\rule[-0.200pt]{0.400pt}{173.689pt}}
\put(219.0,855.0){\rule[-0.200pt]{293.175pt}{0.400pt}}
\put(-22,494){\makebox(0,0){\shortstack{$Q^2 F_{\gamma\pi} (Q^2)$\\ \\ (GeV)}}}
\put(827,10){\makebox(0,0){$Q^2$ (GeV$^2$)}}
\put(219.0,134.0){\rule[-0.200pt]{0.400pt}{173.689pt}}
\put(419,483){\circle*{12}}
\put(450,471){\circle*{12}}
\put(475,532){\circle*{12}}
\put(499,500){\circle*{12}}
\put(523,523){\circle*{12}}
\put(548,569){\circle*{12}}
\put(577,529){\circle*{12}}
\put(619,552){\circle*{12}}
\put(674,515){\circle*{12}}
\put(735,520){\circle*{12}}
\put(796,578){\circle*{12}}
\put(857,552){\circle*{12}}
\put(918,581){\circle*{12}}
\put(1006,561){\circle*{12}}
\put(1180,616){\circle*{12}}
\put(419.0,451.0){\rule[-0.200pt]{0.400pt}{15.418pt}}
\put(409.0,451.0){\rule[-0.200pt]{4.818pt}{0.400pt}}
\put(409.0,515.0){\rule[-0.200pt]{4.818pt}{0.400pt}}
\put(450.0,443.0){\rule[-0.200pt]{0.400pt}{13.731pt}}
\put(440.0,443.0){\rule[-0.200pt]{4.818pt}{0.400pt}}
\put(440.0,500.0){\rule[-0.200pt]{4.818pt}{0.400pt}}
\put(475.0,500.0){\rule[-0.200pt]{0.400pt}{15.418pt}}
\put(465.0,500.0){\rule[-0.200pt]{4.818pt}{0.400pt}}
\put(465.0,564.0){\rule[-0.200pt]{4.818pt}{0.400pt}}
\put(499.0,466.0){\rule[-0.200pt]{0.400pt}{16.622pt}}
\put(489.0,466.0){\rule[-0.200pt]{4.818pt}{0.400pt}}
\put(489.0,535.0){\rule[-0.200pt]{4.818pt}{0.400pt}}
\put(523.0,486.0){\rule[-0.200pt]{0.400pt}{18.067pt}}
\put(513.0,486.0){\rule[-0.200pt]{4.818pt}{0.400pt}}
\put(513.0,561.0){\rule[-0.200pt]{4.818pt}{0.400pt}}
\put(548.0,526.0){\rule[-0.200pt]{0.400pt}{20.958pt}}
\put(538.0,526.0){\rule[-0.200pt]{4.818pt}{0.400pt}}
\put(538.0,613.0){\rule[-0.200pt]{4.818pt}{0.400pt}}
\put(577.0,486.0){\rule[-0.200pt]{0.400pt}{20.717pt}}
\put(567.0,486.0){\rule[-0.200pt]{4.818pt}{0.400pt}}
\put(567.0,572.0){\rule[-0.200pt]{4.818pt}{0.400pt}}
\put(619.0,506.0){\rule[-0.200pt]{0.400pt}{22.163pt}}
\put(609.0,506.0){\rule[-0.200pt]{4.818pt}{0.400pt}}
\put(609.0,598.0){\rule[-0.200pt]{4.818pt}{0.400pt}}
\put(674.0,466.0){\rule[-0.200pt]{0.400pt}{23.608pt}}
\put(664.0,466.0){\rule[-0.200pt]{4.818pt}{0.400pt}}
\put(664.0,564.0){\rule[-0.200pt]{4.818pt}{0.400pt}}
\put(735.0,469.0){\rule[-0.200pt]{0.400pt}{24.813pt}}
\put(725.0,469.0){\rule[-0.200pt]{4.818pt}{0.400pt}}
\put(725.0,572.0){\rule[-0.200pt]{4.818pt}{0.400pt}}
\put(796.0,518.0){\rule[-0.200pt]{0.400pt}{29.149pt}}
\put(786.0,518.0){\rule[-0.200pt]{4.818pt}{0.400pt}}
\put(786.0,639.0){\rule[-0.200pt]{4.818pt}{0.400pt}}
\put(857.0,489.0){\rule[-0.200pt]{0.400pt}{30.594pt}}
\put(847.0,489.0){\rule[-0.200pt]{4.818pt}{0.400pt}}
\put(847.0,616.0){\rule[-0.200pt]{4.818pt}{0.400pt}}
\put(918.0,506.0){\rule[-0.200pt]{0.400pt}{36.135pt}}
\put(908.0,506.0){\rule[-0.200pt]{4.818pt}{0.400pt}}
\put(908.0,656.0){\rule[-0.200pt]{4.818pt}{0.400pt}}
\put(1006.0,492.0){\rule[-0.200pt]{0.400pt}{33.244pt}}
\put(996.0,492.0){\rule[-0.200pt]{4.818pt}{0.400pt}}
\put(996.0,630.0){\rule[-0.200pt]{4.818pt}{0.400pt}}
\put(1180.0,532.0){\rule[-0.200pt]{0.400pt}{40.230pt}}
\put(1170.0,532.0){\rule[-0.200pt]{4.818pt}{0.400pt}}
\put(1170.0,699.0){\rule[-0.200pt]{4.818pt}{0.400pt}}
\sbox{\plotpoint}{\rule[-0.500pt]{1.000pt}{1.000pt}}%
\put(341,670){\usebox{\plotpoint}}
\multiput(341,670)(20.756,0.000){52}{\usebox{\plotpoint}}
\put(1418,670){\usebox{\plotpoint}}
\sbox{\plotpoint}{\rule[-0.200pt]{0.400pt}{0.400pt}}%
\put(341,561){\usebox{\plotpoint}}
\put(430,560.67){\rule{5.300pt}{0.400pt}}
\multiput(430.00,560.17)(11.000,1.000){2}{\rule{2.650pt}{0.400pt}}
\put(341.0,561.0){\rule[-0.200pt]{21.440pt}{0.400pt}}
\put(587,561.67){\rule{5.300pt}{0.400pt}}
\multiput(587.00,561.17)(11.000,1.000){2}{\rule{2.650pt}{0.400pt}}
\put(452.0,562.0){\rule[-0.200pt]{32.521pt}{0.400pt}}
\put(698,562.67){\rule{5.541pt}{0.400pt}}
\multiput(698.00,562.17)(11.500,1.000){2}{\rule{2.770pt}{0.400pt}}
\put(609.0,563.0){\rule[-0.200pt]{21.440pt}{0.400pt}}
\put(810,563.67){\rule{5.300pt}{0.400pt}}
\multiput(810.00,563.17)(11.000,1.000){2}{\rule{2.650pt}{0.400pt}}
\put(721.0,564.0){\rule[-0.200pt]{21.440pt}{0.400pt}}
\put(944,564.67){\rule{5.541pt}{0.400pt}}
\multiput(944.00,564.17)(11.500,1.000){2}{\rule{2.770pt}{0.400pt}}
\put(832.0,565.0){\rule[-0.200pt]{26.981pt}{0.400pt}}
\put(1056,565.67){\rule{5.300pt}{0.400pt}}
\multiput(1056.00,565.17)(11.000,1.000){2}{\rule{2.650pt}{0.400pt}}
\put(967.0,566.0){\rule[-0.200pt]{21.440pt}{0.400pt}}
\put(1190,566.67){\rule{5.300pt}{0.400pt}}
\multiput(1190.00,566.17)(11.000,1.000){2}{\rule{2.650pt}{0.400pt}}
\put(1078.0,567.0){\rule[-0.200pt]{26.981pt}{0.400pt}}
\put(1324,567.67){\rule{5.541pt}{0.400pt}}
\multiput(1324.00,567.17)(11.500,1.000){2}{\rule{2.770pt}{0.400pt}}
\put(1212.0,568.0){\rule[-0.200pt]{26.981pt}{0.400pt}}
\put(1347.0,569.0){\rule[-0.200pt]{21.440pt}{0.400pt}}
\end{picture}}
\end{center}
\caption[*]{The $\gamma\rightarrow\pi^0$ transition form factor.  The
solid line is the full prediction including the QCD correction
[Eq. (\ref{fgampi})]; the dotted line is the LO prediction
$Q^2F_{\gamma\pi}(Q^2) = 2f_\pi$.}
\label{fgammapi}
\end{figure}
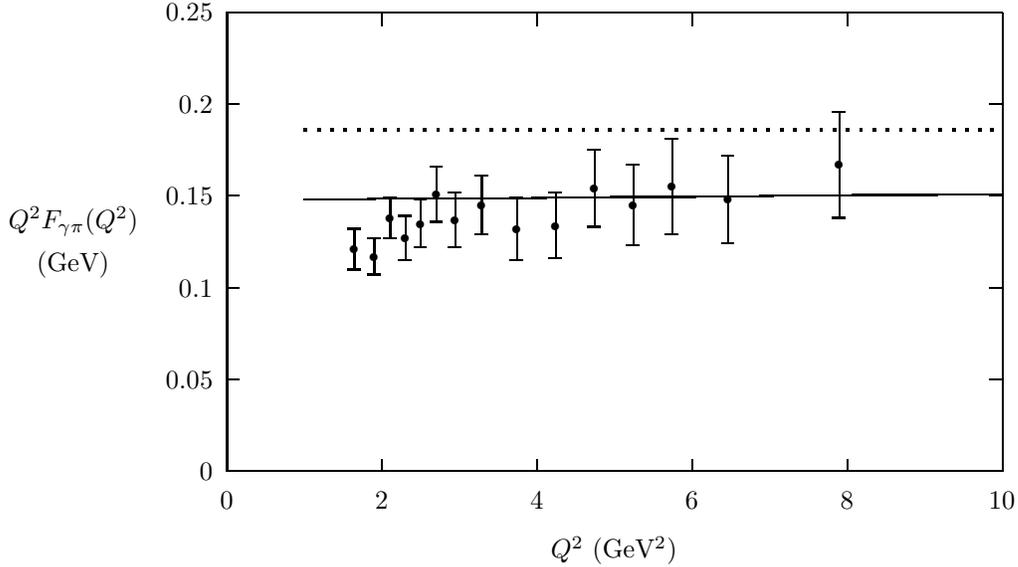

In Fig. \ref{fgammapi}, we compare the recent CLEO
data \cite{Dominick} for the photon to pion transition form factor with
the prediction
\begin{equation}
Q^2 F_{\gamma\pi}(Q^2)= 2 f_\pi \left( 1
- {5\over3} {\alpha_V(e^{-3/2} Q)\over \pi}\right).
\label{fgampi}
\end{equation}
The 
flat scaling of the $Q^2 F_{\gamma \pi}(Q^2)$ data from $Q^2 = 2$
to $Q^2 = 8$ GeV$^2$ provides an important confirmation of the
applicability of leading twist QCD to this process. The magnitude of
$Q^2 F_{\gamma \pi}(Q^2)$ is remarkably consistent with the predicted
form, assuming the asymptotic distribution amplitude and including the
LO QCD radiative correction with $\alpha_V(e^{-3/2} Q)/\pi \simeq
0.12$.  Radyushkin,\cite {Radyushkin} Ong \cite{Ong} and Kroll \cite
{Kroll} have also noted that the scaling and normalization of the
photon-to-pion transition form factor tends to favor the asymptotic
form for the pion distribution amplitude and rules out broader
distributions such as the two-humped form suggested by QCD sum rules.
\cite{CZ}   One cannot obtain a unique solution for the
nonperturbative wavefunction from the $F_{\pi\gamma}$ data alone.
However, we have the constraint that
\begin{equation}
{1\over 3}\langle {1\over 1-x}\rangle \left[ 1-{5\over
3}{\alpha_V(Q^*)\over\pi} \right]\simeq 0.8
\end{equation}
(assuming the renormalization scale we have chosen in
Eq. (\ref{qsquaretransition}) is approximately correct).  Thus one
could allow for some broadening of the distribution amplitude with a
corresponding increase in the value of $\alpha_V$ at low scales.

\begin{figure}[htbp]
\begin{center}
\hbox to 3in{\input{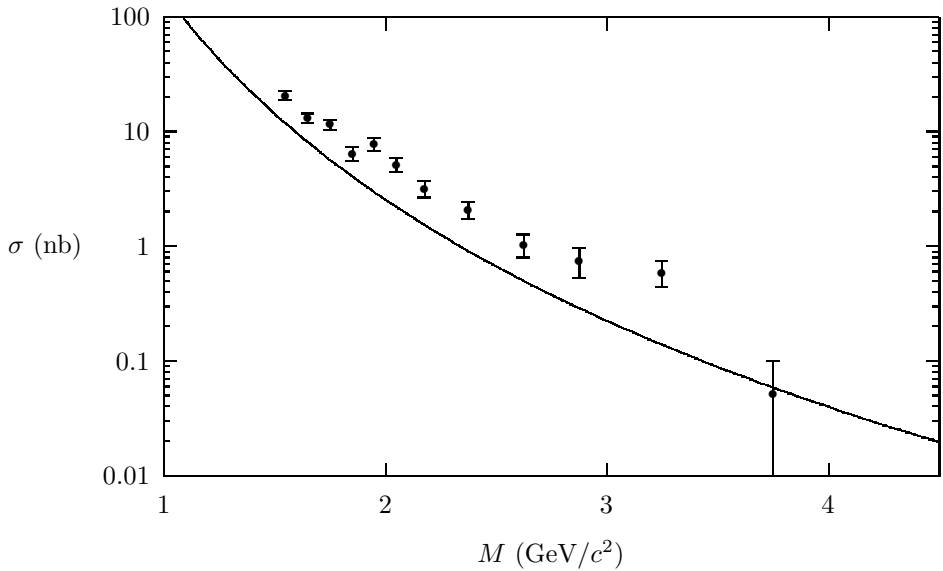}}
\end{center}
\caption[*]{Two-photon annihilation cross section $\sigma(\gamma \gamma
\to\pi^+\pi^-,K^+K^-)$ as a function of CMS energy, for
$|\cos\theta^*|<0.6$.}
\label{sigma}
\end{figure}

We have also analyzed the $\gamma\gamma \rightarrow \pi^+\pi^-, K^+
K^-$ data. These data exhibit true leading-twist scaling (Fig.
\ref{sigma}), so that one would expect this process to be a good test
of theory. One can show that to LO
\begin{equation}
{{d\sigma\over dt}\left(\gamma\gamma\rightarrow\pi^+\pi^-\right) \over
{d\sigma\over dt}\left(\gamma\gamma\rightarrow\mu^+\mu^-\right)} =
{4|F_\pi(s)|^2\over 1-\cos^4\theta_{c.m.}}
\end{equation}
in the CMS, where $dt=(s/2) d(\cos\theta_{c.m.})$ and here $F_\pi(s)$
is the {\em time-like} pion form factor.  The ratio of the time-like
to space-like pion form factor for the asymptotic distribution
amplitude is given by
\begin{equation}
{|F^{(\rm timelike)}_\pi(-Q^2)|\over F^{(\rm spacelike)}_\pi(Q^2)}
= {|\alpha_V(-Q^{*2})|\over \alpha_V(Q^{*2})}.
\label{ratio}
\end{equation}
If we simply continue Eq. (\ref{frozencoupling}) to negative values of
$Q^2$ then for $1 < Q^2 < 10$ GeV$^2$, and
hence $0.05 < Q^{*2} < 0.5$ GeV$^2$, the ratio of couplings in
Eq. (\ref{ratio}) is of order 1.5.  Of course this assumes the
analytic application of Eq. (\ref{frozencoupling}).  Thus if we assume
the asymptotic form for the distribution amplitude, then we predict
$F^{(\rm timelike)}_\pi(-Q^2) \simeq (0.3~{\rm GeV}^2)/Q^2$ and hence
\begin{equation}
{{d\sigma\over dt}\left(\gamma\gamma\rightarrow\pi^+\pi^-\right) \over
{d\sigma\over dt}\left(\gamma\gamma\rightarrow\mu^+\mu^-\right)
}\simeq {.36 \over s^2} {1\over 1-\cos^4\theta_{c.m.}}.
\label{twophotonratio}
\end{equation}
The resulting prediction for the combined cross section
$\sigma(\gamma\gamma\to\pi^+\pi^-, K^+K^-)$\footnote{The contribution
from kaons is obtained at this order simply by rescaling the
prediction for pions by a factor $(f_K/f_\pi)^4\simeq 2.2$.} is shown
in Fig. \ref{sigma}, along with CLEO data.\cite{Dominick}
Considering the possible contribution of the resonance $f_2(1270)$,
the agreement is reasonable.

\vspace{.5cm}

We also note that the normalization of $\alpha_V$ could be larger at
low momentum than our estimate. This would also imply a broadening of
the pion distribution amplitude compared to its asymptotic form since
one needs to raise the expectation value of $1/(1-x)$ in order to
maintain consistency with the magnitude of the $F_{\gamma \pi}(Q^2)$
data. A full analysis will then also require consideration of the
breaking of scaling from the evolution of the distribution amplitude.
In any case, we find no compelling argument for significant
higher-twist contributions in the few GeV regime from the hard
scattering amplitude or the endpoint regions, since such corrections
violate the observed scaling behavior of the data.

The analysis we have presented here suggests a systematic program for
estimating exclusive amplitudes in QCD (including exclusive $B$-decays)
which involve hard scattering.  The central input is $\alpha_V(0)$, or
\begin{equation}
\overline{\alpha_V} = {1\over{Q_0^2}}\int_0^{Q_0^2}d{Q^\prime}^2
\alpha_V({Q^\prime}^2),\;\; Q_0^2 \leq 1\;\;{\rm GeV}^2,
\label{old21}
\end{equation}
which largely controls the magnitude of the underlying quark-gluon
subprocesses for hard processes in the few-GeV region.  In this work,
the mean coupling value for $Q_0^2 \simeq 0.5$ GeV$^2$ is
$\overline{\alpha_V} \simeq 0.38.$ The main focus will then be to
determine the shapes and normalization of the process-independent meson
and baryon distribution amplitudes.

The leading-twist scaling of the observed cross sections for exclusive
two-photon processes and other fixed $\theta_{cm}$ reactions can be
understood if the effective coupling $\alpha_V(Q^*)$ is approximately
constant in the domain of $Q^*$ relevant to the underlying hard
scattering amplitudes.  In addition, the Sudakov suppression of the
long-distance contributions is strengthened if the coupling is frozen
because of the exponentiation of a double log series.    We have also
found that the commensurate scale relation connecting the heavy quark
potential, as determined from lattice gauge theory, to the
photon-to-pion transition form factor is in excellent agreement with
$\gamma e \to \pi^0 e$ data assuming that the pion distribution
amplitude is close to its asymptotic form $\sqrt{3}f_\pi x(1-x)$.  We
also reproduce the scaling and approximate normalization of the
$\gamma\gamma \rightarrow \pi^+\pi^-, K^+ K^-$ data at large momentum
transfer. However, the normalization of the space-like pion form factor
$F_\pi(Q^2)$ obtained from electroproduction experiments is somewhat
higher than that predicted by the corresponding commensurate scale
relation. This discrepancy may be due to systematic errors introduced
by the extrapolation of the $\gamma^* p \to \pi^+ n$ electroproduction
data to the pion pole.

\section {Acknowledgments}

Much of the content of these lectures is based on 
collaborations with  Matthias Burkardt, Sid Drell, 
Paul Hoyer, Chueng Ji, Marek Karliner, Peter Lepage, Hung
Jung Lu, Bo-Qiang Ma,  Alex Pang,  Hans Christian Pauli, Dave Robertson, 
Ivan Schmidt, Felix Schlumpf, and
Ramona Vogt.  I am particularly
grateful to Professors Chueng Ji and Dong-Pil Min for their 
invitation to this
school. This work is supported in part by the U.S. Department of Energy under
contract no. DE--AC03--76SF00515.

\end{document}